\documentclass[conference]{IEEEtran}
\IEEEoverridecommandlockouts
\usepackage{cite}
\usepackage{amsmath,amssymb,amsfonts}
\usepackage[linesnumbered,ruled,lined]{algorithm2e}
\usepackage{graphicx}
\usepackage[caption=false]{subfig}
\usepackage{textcomp}
\usepackage{xcolor}
\usepackage{url}

\usepackage{mathptmx}
\usepackage[T1]{fontenc}
\def\BibTeX{{\rm B\kern-.05em{\sc i\kern-.025em b}\kern-.08em
   T\kern-.1667em\lower.7ex\hbox{E}\kern-.125emX}}

\usepackage{booktabs}
\usepackage{pifont}
\usepackage{array}
\usepackage{enumitem}
\newcolumntype{L}{>{\centering\arraybackslash}m{3cm}}

\graphicspath{{/}}

\usepackage{fancyhdr}

\usepackage{multirow}

\begin{document}

\pagestyle{fancy}
%... then configure it.
\fancyhead{} % clear all header fields
\fancyhead[R]{\textbf{Preprint version}}
\fancyfoot{} % clear all footer fields
% \fancyfoot[LE,RO]{\thepage}

\title{Physics-informed Graphical Neural Network for Power System State Estimation \\
}

\author{\IEEEauthorblockN{Quang-Ha Ngo}
\IEEEauthorblockA{\textit{Clarkson University} \\
hanq@clarkson.edu}
\and
\IEEEauthorblockN{Bang L. H. Nguyen}
\IEEEauthorblockA{\textit{Los Alamos National Lab} \\
bangnguyen@ieee.org}
\and
\IEEEauthorblockN{Tuyen V. Vu}
\IEEEauthorblockA{\textit{INS Engineering} \\
tvu@ins-engineering.com}

\and
\IEEEauthorblockN{Jianhua Zhang}
\IEEEauthorblockA{\textit{Clarkson University} \\
jzhang@clarkson.edu}

\and
\IEEEauthorblockN{Tuan Ngo}
\IEEEauthorblockA{\textit{INS Engineering} \\
tngo@ins-engineering.com}

}

\maketitle

\begin{abstract}

    State estimation is highly critical for accurately observing the dynamic behavior of the power grids and minimizing risks from cyber threats. However, existing state estimation methods encounter challenges in accurately capturing power system dynamics, primarily because of limitations in encoding the grid topology and sparse measurements. This paper proposes a physics-informed graphical learning state estimation method to address these limitations by leveraging both domain physical knowledge and a graph neural network (GNN). We employ a GNN architecture that can handle the graph-structured data of power systems more effectively than traditional data-driven methods. The physics-based knowledge is constructed from the branch current formulation, making the approach adaptable to both transmission and distribution systems. The validation results of three IEEE test systems show that the proposed method can achieve lower mean square error more than 20$\%$ than the conventional methods.

\end{abstract}


\section{Introduction}
State estimation is widely recognized as an essential part of energy management, as it improves the observability and operation of smart grids. The performance of state estimation relies heavily on the measurement data obtained from Phasor Measurement Units (PMUs) and Supervisory Control and Data Acquisition Systems (SCADA). By utilizing these sources of information, state estimation can effectively provide real-time system status, such as voltage amplitude and phase angle to ensure the stability and reliability of grid operations.

In recent years, smart grids have been developing remarkably with the growing deployment of distributed renewable generators, power electronic devices and large-scale interconnection\cite{IEA2022}. This deployment challenges the modern power systems by amplifying rapid voltage fluctuations in the operational conditions. The sudden voltage changes significantly degrade the equipment performance and cause system instability \cite{7150418}. Consequently, real-time power system monitoring becomes more critical, not only for anomaly detection and protection, but also for energy control and management. However, conventional state estimation techniques are struggling to keep up with the increasing complexity and scalability of the smart power grids \cite{8466598}. The demand for more accurate system situational awareness, coupled with the limitations of traditional techniques, has motivated the development of a new approach for power system state estimation.

 Existing state estimation efforts for power systems can be categorized into model-based and machine learning based approaches \cite{8466598, zhao2019power, zamzam_data-driven_2019, azimian_state_2022}. In the domain of model-based state estimation, two directions have emerged as key areas of focus: (1) static state estimation (SSE) and (2) dynamic state estimation (DSE) \cite{vu_cyber-physical_2020}. SSE determines the unknown states of a power system by applying the weighted least squares (WLS) method to solve power flow equations. The WLS process only concerns the snapshot measurement samples so SSE lacks the capability of capturing the time-varying
dynamics and handling uncertainties in a power system \cite{zhao2019power}. Given this limitation, DSE with Kalman filter techniques, such as Extended Kalman Filter (EKF), Unscented Kalman Filter (UKF) \cite{zhao_robust_2017} is proposed. In \cite{nguyen2021distributed}, the distributed DSE estimation, which effectively applies the Kalman filter method to the physical equations of branch currents for a power system, significantly outperforms SSE. However, power systems are always facing limited available measurements, hindering the ability of model-based schemes including SSE and DSE to fully estimate states. To this end, research groups have increasingly redirected their focus towards data-driven approaches\cite{s21062085}.
 %%%

\setlength{\tabcolsep}{0.35\tabcolsep}
\begin{table*}[ht]
%\centering

\caption{Summary of Technical Differences With Previous PINN-based Literature}
\begin{tabular}{l ccccc ccccc}
\toprule
\textbf{Aspects}
& AS et al. \cite{9072507}
& Lei et al. \cite{9126036}
& Di et al. \cite{10143253}
& Tong et al. \cite{10238825}
& Wang et al. \cite{9647978}
& Lau et al. \cite{pagnier2021physics}
& Jon et al. \cite{ostrometzky2019physics}
& Jongh et al.\cite{de2022physics}
& Ours\\
\cmidrule(lr){1-1} \cmidrule(lr){2-2} \cmidrule(lr){3-3} \cmidrule(lr){4-4} \cmidrule(lr){5-5} \cmidrule(lr){6-6} \cmidrule(lr){7-7} \cmidrule(lr){8-8} \cmidrule(lr){9-9} \cmidrule(lr){10-10}

Small datasets & 
\centering \ding{55} &\ding{55} & \ding{51}  & \ding{55} & \ding{51} & \ding{51} & \ding{55} & \ding{55} & \ding{51} &  \\

\midrule
Dynamic estimation & \centering \ding{55} &\ding{51} & \ding{51}  & \ding{55} & \ding{55} & \ding{51} & \ding{51} & \ding{55} & \ding{51} &  \\

\midrule
 System physics \\ integration & \centering \ding{55} &\ding{55} & \ding{55}  & \ding{55} & \ding{51} & \ding{51} & \ding{51} & \ding{51} & \ding{51} & \\

\midrule
Scalability for \\ large systems& \centering \ding{51} &\ding{55} & \ding{51}  & \ding{51} & \ding{55} & \ding{55} & \ding{55} &  \ding{55} & \ding{51} &\\

\midrule
Performance on \\unbalanced systems& \centering \ding{51} &\ding{55} & \ding{51}  & \ding{51} & \ding{51} & \ding{55} & \ding{51} &  \ding{55} & \ding{51} &\\

% \midrule
% Fuzzy logic- \\based fusion & \centering \ding{55} &\ding{55} & \ding{55}  & \ding{55} & \ding{55} & \ding{55} & \ding{55} & \ding{55} & \ding{51} &  \\

\bottomrule
\label{table:existingPINN}
\end{tabular}
\end{table*}

%%%
Data-driven methods are designed to effectively establish the functional relationships between measurements and system states without requiring physical models. Establishing this relationship enables significant advancements in state estimation by leveraging the large number of data during the training process \cite{s21062085}. Therefore, compared to model-based methods, the data-driven approach can excel in situations with complex or poorly understood system dynamics and uncertainties. Generally, recent models as the data-driven approaches include multilayer perceptron (MLP), recurrent neural networks (RNN), long short term memory (LSTM), convolutional neural networks (CNN), generative adversarial networks (GAN), and graph neural network (GNN) \cite{9265470, yarlagadda_power_2021, he_power_2020, zhang_real-time_2019}. Among these models, GNN emerged as a prominent data-driven approach since 2020, demonstrating significant advancements in handling datasets represented as graphs. To understand why this deep learning model can outperform other data-driven methods by encoding both graph structure and node features, a small number of studies have been carried out. Interestingly, all of these have investigated graph network potential for effective capture of spatio-temporal correlations in the context of regression and classification tasks \cite{zhou2020graph}. GNN, with its message-passing mechanism, can definitely enhance precision and robustness for power system state estimation because a dynamic power system, consisting of vertices, and edges, exhibits a structure the same as a graph. In \cite{10050763, Park2023DistributedPS, KUNDACINA2023101056, nguyen_spatial-temporal_2022}, studies have proposed the effective use of GNN for state estimation by exploiting attributes of graph structures. However to be accurate, these pure data-reliant models require vast amounts of data, which is unrealistic in power systems because data collection is restricted due to constraints on resources, sampling rate and accuracy. A more holistic concern is that the data-driven approach lacks the consideration for the underlying physical aspects of the system, leading to unsatisfactory results\cite{9429985}. To this end, guaranteeing realistic and effective situational awareness requires a rethinking of the state estimation framework.

As a feasible solution to resolve these challenges, a physics-informed neural network (PINN) has recently emerged as a hybrid paradigm to enhance estimation capability. Firstly, physics-informed methods are designed to incorporate the known physical principles such as Kirchhoff's circuit laws, as well as constraints governing power systems as a part of their modeling framework \cite{RAISSI2019686}. Therefore, they can favorably enable the utilization of domain-specific knowledge for more accurate and reliable state estimation. In 2019, a study in \cite{misyris2020physics} showcased the pioneering integration of mathematical formulations describing the system’s behavior into machine learning techniques with a single-machine infinite bus system. Secondly, physics-informed approaches provide enhanced accuracy, reduced computational costs, and lower data requirements compared to existing machine learning techniques \cite{9743327}. The fundamental concept of PINN has been initially examined and explored in different engineering fields, including chemistry \cite{ji2021stiff}, power electronics \cite{9779551}, etc. Several studies have also investigated the integration of physics into data-driven approaches, particularly through GNN, seeking to understand potential enhancements in state estimation. 

The summary of major technical differences between our proposed method with existing PINN-based models is outlined in Table \ref{table:existingPINN}. In \cite{9072507}, authors focus on reconstructing the admittance matrix to reduce the number of neural network parameters. Unlike physics-informed machine learning, this approach does not consider incorporating domain knowledge into the learning process. Works in \cite{10143253, 10238825} utilize variants of GNN, which assign importance weights to neighboring nodes for the physical information but lack adaptability to topology changes since they do not leverage power system models. The paper \cite{9647978} proposes a PINN that leverages GNN and power flow calculations to iteratively update the GNN transition function for state estimation but this method has computational challenges scaling to large systems. \cite{pagnier2021physics} combines model-based approaches derived from power flow equations and GNN to perform dynamic state estimation but their approach only investigates transmission systems without exploring distribution systems. \cite{ostrometzky2019physics} and \cite{de2022physics} incorporate physical information for physics-informed state estimation by leveraging power flow calculations. However, their approaches only focus on steady-state analysis, limiting power system dynamics. In summary, existing state estimation methods have notable limitations including:
(1) \textit{Limited observability}, (2) \textit{Neglecting physical aspects}, (3) \textit{Scalability challenge}, (4) \textit{Dynamic estimation gap}.

To address these limitations, this paper develops a novel physics-informed graph learning method for power distribution and transmission systems. The objective is to estimate bus voltage and phase angles under load-change scenarios. The main contributions can be summarized as follows:

\begin{itemize}
  \item We propose the physics-informed graphical learning approach that combines the model-based dynamic state estimation and the temporal-spatial graphical neural network.
  
  \item The model-based dynamic state estimation utilizes the Kalman filter is applied for $\mu$PMU measurements with high accuracy. The graphical neural network yield estimated states of the entire system with both $\mu$PMU measurements and traditional measurements of voltage amplitude, active and reactive powers.
  \item The graphical neural network considers the data under both temporal and spatial correlations. With the combination of the model-based Kalman filtering, the training process of the GNN model can be enhanced since the model-based estimation contains the physical information of the system.

  \item Comprehensive case studies are performed with IEEE 5-bus, 123-bus and 8500-bus systems.
\end{itemize}

The remaining sections are structured as follows. In Section II, the paper outlines the methodology encompassing both the state-space equations for model-based estimation and GNN as a data-driven method. Additionally, this section gives background on the concept of PINN. Section III presents the implementation of physics-informed state estimation, including data preparation and model training. Section IV describes case studies and evaluates the effectiveness of our state estimation scheme. Finally, in Section V, we present our conclusions.

% \cite{boyaci_graph_2022}, saxena_agent-based_2020 \cite{mestav_bayesian_2019, massignan_bayesian_2022}, \cite{10131654}, \cite{raissi2020hidden}, nguyen20221,\cite{nguyen_spatial-temporal_2022}, 

% ,\cite{nguyen2021distributed},\cite{zhao2019power},
% \cite{yu2017spatio}, 
% \cite{xiong2022fused},\cite{9779551},\cite{9429985}, 
% \cite{misyris2020physics}, 
% \cite{9743327}, 
% \cite{kipf2017semisupervised}, 
% \cite{azimian_state_2022},\cite{tian_neural-network-based_2021},\cite{labach2019survey},\cite{fey2019fast},\cite{he_power_2020},  \cite{zhang_real-time_2019}, 

%\cite{ mao_multiarea_2022},

% \textcolor{red}{A paragraph/section to highlight urgency of developing new approach for power system state estimation. Why has the current approach not applicable? What are the challenges and what has been done.? What has been suggested in literature for the new approach to address these challenges specific to power system state estimation?}\\

% \textcolor{red}{How might GNN solve the limitations of the other methods? what are the special features of power systems that need
% GNN?What is the limitation of GNN (purely Data-Driven methods)? What has been proposed?}
% PIML, PINN
% \textcolor{red}{Reasearch approach: Physics-Inform Graph Learning for State Estimation. Why Physics Integration is natural and effective in State Estimation. Are we the first people to apply PINN for SE. How is your PINN different from the PINN or hybrid data-driven approach in other papers (cite)? Concentrate on the research gap and explain why our innovation might work}

%\vspace{-1.5mm}
\section{Methodology}
This section outlines three key components of the proposed state estimation method: a model-based approach, GNN, and an innovative fusion that combines the strengths of both techniques through a physics-informed paradigm.

\subsection{Model-Based Estimation of Dynamic Power Systems}
\vspace{-3pt}
A power system typically include generators, loads, buses, and the interconnections between them. In order to effectively estimate the dynamic behavior of power systems, the differential equations of currents between buses \cite{nguyen2021distributed} can be used within the $abc$-frame as: 
\begin{align} \label{eq:MB1}
&\frac{di_{ij,\;abc}}{dt} = -\frac{R_{ij}}{L_{ij}}i_{ij,\;abc} \;+ \frac{1}{L_{ij}}(v_{i,\; abc} \;- v_{j,\;abc})
%
%& v_{i, k+1} = f(v_{i,k}, i_{i,k}) \; + w_{v,k}
\end{align}
Equation (\ref{eq:MB1}) illustrates the relationship between the branch current flow denoted as $i_{ij}$, which occurs between buses $i$ and $j$, and the corresponding bus voltages $v_i$ and $v_j$. The line resistance $R_{ij}$ and inductance $L_{ij}$ represent the resistance and inductance of the line connecting the buses. In cases where bus voltages of the dynamic model are unavailable, these voltages are considered unknown inputs. The line parameters are assumed to be fixed and negligible with shunt capacitance on the dynamic behavior of micro-grids.

The voltage and branch current measurement functions can be expressed as follows:
\vspace{1mm}
\begin{align} \label{eq:MB2}
&\left[ \begin{array}{c} z_{vi, abc} \\ z_{cij, abc} \end{array} \right] = \begin{bmatrix} I & 0 \\ 0 & I \end{bmatrix} . \left[ \begin{array}{c} v_{i,abc} \\ i_{ij,abc} \end{array}  \right] \:+ \left[ \begin{array}{c} e_{vi} \\ e_{cij} \end{array}  \right]
\end{align}
\vspace{1mm}
where $I$ is the identity matrix. $z_{vi}$ and $z_{cij}$ represent the measurements of bus voltage at bus $i$ and branch current between bus $i$ and $j$, respectively, and $e_{vi}$ and $e_{cij}$ denote the corresponding measurement noises. 

Based on the  differential branch current and measurement equations, state estimation problem can be formulated as the state-space representation below.
\begin{align} \label{eq:MB3}
& \begin{cases} x_k = Ax_{k-1} + Bu_{k-1} + w_{k-1},\\ z_{x,k} = Cx_k + v_{x,k}, \\ z_{u,k} = Du_k + v_{u,k} \end{cases}
\end{align}
In the given formulation, $x_k$ represents the state vector at time step $k$ and $u_k$ denotes the unknown input vector. The state and input matrices are represented by $A$ and $B$, respectively, while $C$ and $D$ represent the measurement matrices. The modeling errors are denoted as $w_{k-1}$. The state and input measurements, denoted by $z_{x,k}$ and $z_{u,k}$, are accompanied by the measurement noises $v_{x,k}$ and $v_{u,k}$. 

From (\ref{eq:MB3}), the relationship between the system's input and measurement can be calculated as
\begin{align} \label{eq:MB4}
& z_{x,k} = CAx_{k-1} + CBu_{k-1} + Cw_{k-1} + v_{x,k} 
\end{align}
The estimates $\begin{bmatrix} \hat{x}_{k-1} & \hat{u}_{k-1} \end{bmatrix} $ and the covariance matrix $P_{k-1}$ results from solving Equation (\ref{eq:MB4}) along with the measurements at $(k-1)$. After that, the predicted value and covariance can be obtained with
\begin{subequations}
\begin{align} 
\label{eq:MB5}
&\hat{x}_{k|k-1} = A\hat{x}_{k-1} + B\hat{u}_{k-1}, \\
\label{eq:MB6}
& P_{x,\:k|k-1} = \begin{bmatrix} A & B \end{bmatrix} P_{k-1} \begin{bmatrix} A \\ B \end{bmatrix} +Q_{k-1}
\end{align}
\end{subequations}
As a result, the updated estimates can be achieved by
\begin{subequations}
\begin{align}
\label{eq:MB7}
&\hat{x}_k = \hat{x}_{k|k-1} + K_k(z_{x,k}-C\hat{x}_{k|k-1}), \\
\label{eq:MB8}
& P_{x,k} = (I-K_kC)P_{k|k-1}
\end{align}
\end{subequations}
where $K_k$ represents the optimal Kalman gain, and the covariance matrix is denoted as $Q_{k-1}$. 
% \textcolor{red}{qualify all equations}

%\vspace{2pt}
Bang \cite{nguyen2021distributed} introduced a distributed dynamic state-input estimation algorithm in the $dq$ frame. This algorithm offers improved performance compared to traditional WLS method, as it utilizes the Kalman-based filtering algorithm with unknown inputs. Inspired by the Bang's idea, our model-based estimation approach can be performed for the $abc$ frame. This estimation procedure is outlined in Algorithm \ref{alg:two}. By applying this algorithm using available measurements from PMUs, the voltages on a power system can be observed with a high accuracy. However, model-based state estimation relies on the availability of measurements and inputs to accurately estimate the system state. If several voltages in the system are not inherently measured, it can make the estimation intractable. 
we perform the model-based dynamic state estimation only in areas with $\mu$PMU installed, which provide us the voltage and currents measured with time synchronization, and then combine with GNN for state estimation in the entire systems.

\begin{figure}[t!]
    \centering
    \includegraphics[height = 2.0cm, width=6.4cm]{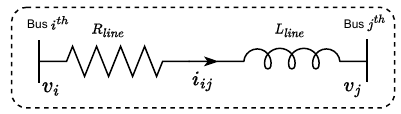}
    \vspace{-3mm}
    \caption{The power line in a distribution system}
    \label{fig:transmission_line}
    \vspace{-3mm}
\end{figure}

\RestyleAlgo{ruled}
%% This is needed if you want to add comments in
%% your algorithm with \Comment
\SetKwComment{Comment}{/* }{ */}
\begin{algorithm}[b!]

\caption{Model-Based State Estimation}\label{alg:two}
\KwData{$z_{x,k-1}, z_{u,k-1}, z_{x,k}$}
\KwResult{$x_{k}$}
Declare line parameters and observation noises.\\
Initialize state vectors and covariance matrices.\\
\textbf{Repeat}\\
\hspace{0.25cm} Solve (\ref{eq:MB4}) for $\hat{x}_{k-1}, \hat{u}_{k-1}$ and $P_{k-1}$.\\
\hspace{0.25cm} Predict $\hat{x}_{k|k-1}$ and $P_{x,k|k-1}$ using (\ref{eq:MB5}) and (\ref{eq:MB6}).\\
\hspace{0.25cm} Update $\hat{x}_k$ and $P_{x,k}$ using (\ref{eq:MB7}) and (\ref{eq:MB8}).\\
\textbf{End}\\
Return estimation results as voltage sinusoidal signals.

\end{algorithm}
%\vspace{-1mm}
\subsection{Graph Neural Network Structure}
\begin{figure}[tbp]
    \vspace{-2mm}
    \centering
    \includegraphics[height = 2.6cm, width=5.8cm]{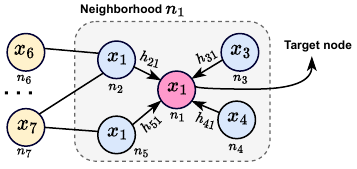}
    \vspace{-2mm}
    \caption{A message-passing process in each GNN layer}
    %\vspace{-2mm}
    \label{fig:GNN_message_passing}
    \vspace{-1mm}
\vspace{-3mm}
\end{figure}

 The primary objective of GNN is to effectively generalize the relationships between  the attributes of nodes and/or edges with corresponding outputs\cite{zhou2020graph}. These outputs in graph learning can encompass tasks such as classification or regression. By leveraging input measurements, the state estimation process is formulated within a message-passing framework of GNN for efficient information propagation and inference \cite{9654642}. Given an example of a message-passing scheme illustrated in Fig. \ref{fig:GNN_message_passing}, node $n_{1}$ aggregates information from its neighboring nodes and updates its own representation based on the gathered information, allowing for the learning of complex patterns and relationships in the graph data. Likewise, this applies to every node within the graph. The message denoted as $h$ is exchanged among neighboring nodes to facilitate node-level state estimation. Overall, the message-passing scheme serves as a fundamental operation in each GNN layer.
 
The structure of GNN typically involves multiple layers of node and graph-level operations, enabling the model to capture the spatial-temporal information inherent in bus voltages in power systems. This is achieved by utilizing a localized first-order approximation of spectral graph convolutions \cite{kipf2017semisupervised}. The mathematical formulation of GNN layers can be expressed as

\vspace{-2mm}
\begin{equation} H_{l+1}^{i} = \sigma (\tilde{D}^{-\frac{1}{2}} \tilde{A} \tilde{D}^{-\frac{1}{2}} H_{l}^{i} W_{l}).
\end{equation}
\vspace{-3mm}

Where $\tilde{A} = A + I_{N}$ denotes the adjacency matrix with self-loops, $\tilde{D}$ is the degree matrix derived from the matrix $\tilde{A}$, $I_{N}$ is the identity matrix, and $\sigma$ is a non-linear activation function. The trainable weights $W_{l}$ of layer $l$ are updated iteratively using back-propagation to minimize the loss function. Overall, the GNN architecture will reveal the evolution of the system state in correlation with adjacent states, enhancing the accuracy of the state estimation capabilities.

\vspace{1mm}
\subsection{Physics-Informed Learning on Graphs}
Physics-informed machine learning offers an advantage over both model-based and data-driven approaches. Unlike model-based methods, PINN can estimate states from datasets without being constrained by limitations of system observability. In addition, accounting for model uncertainties is rather simplistic, which likely misses important features for state estimation. Therefore, in practice, model-based methods have either failed to achieve the reliable estimated states or have been vulnerable to cyber attacks. Another concern is that data-driven approaches remain ineffective for state estimation since they primarily focus on Euclidean data, which is not graph-structured \cite{9654642}. Power system data, in contrast, are non-Euclidean, making it hard to define localized convolutional filters as normally seen in the machine learning model like CNN \cite{9395439}. While the geometric learning approach with GNN can address the non-Euclidean power system data and structure, the critical challenge to adopting it into state estimation is the data quality problem. Data can be limited or imbalanced, especially abnormal data often derived from the nature of power transmission or distribution systems, as well as constraints in data collection. These challenges can lead to model inaccuracy due to either over-fitting or sub-optimal solutions. To overcome the data sparsity and accurately regularize the system’s behavior, researchers have explored the integration of partial differential equations from dynamic systems into machine learning models, known as PINNs \cite{9064519}. 

%\subsection{Investigated Power Systems}
\begin{figure}[tbp]
    \vspace{-3mm}
    \centering
    \includegraphics[height = 4.3cm, width=6.4cm]{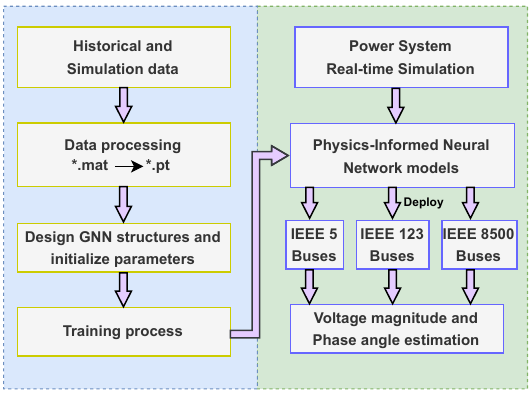}
    \vspace{-2mm}
    \caption{The overall flowchart of the PINN framework}
    %\vspace{-2mm}
    \label{fig:flow_chart}
    \vspace{-2mm}
\vspace{-3mm}
\end{figure}
In the general PINN formulation, a neural network is trained to learn the solution of a partial differential Equation (\ref{eq:PIML1}) by minimizing a loss function that consists of two terms \cite{9779551, MOHAMMADIAN2023109551}. The first term is the machine learning model loss $Loss_u$, which measures the discrepancy between the observed data and the predictions of the neural network. The second term is the physics loss $Loss_f$, which is a residual of the partial differential equation. To incorporate the physics constraints, the neural network is trained to satisfy the governing equations by taking derivatives of the neural network outputs concerning the input variables. These derivatives are then combined with the PDEs to form the physics loss term. By minimizing the overall loss function, the neural network learns to produce solutions that not only effectively fit the provided data but also adhere to the underlying physics of the system \cite{9743327}. Mathematically, the basic formulation of PINN, which include a generic form of partial differential equations along with a function approximation provided by a neural network, can be expressed as Equations (\ref{eq:PIML2}) and (\ref{eq:PIML3}), respectively \cite{9779551}.
\vspace{-1pt}
\begin{subequations}
\begin{align} 
\label{eq:PIML1}
&u_t+N_{x}[u] = 0, x\in \Omega , t \in [0, T].\\
\label{eq:PIML2}
&f:=u_t + N_{x}[u;\lambda].\\
\label{eq:PIML3}
&Loss = Loss_u + Loss_f
\end{align}
\end{subequations}
where $u(x,t)$ refers to the solution of the differential equations with the state variable $x$ and the time $t$. $N$ is a nonlinear differential operator, describing the dynamics of the system.

\begin{figure}[t!]%[htp]
    \vspace{-4mm}
    \centering
    \includegraphics[height = 4.7cm, width=8.4cm]{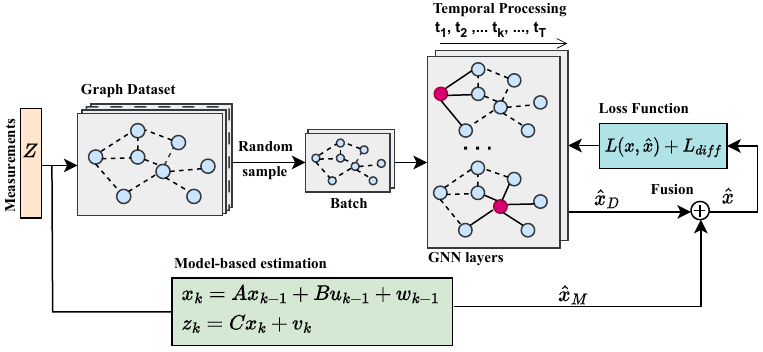}
    \vspace{-3mm}
    \caption{The training process diagram}
    %\vspace{-2mm}
    \label{fig:Training_process}
    \vspace{-4mm}
\end{figure}

 Our proposed method, inspired by the concept of PINN, is depicted in Fig. \ref{fig:Training_process}. Initially, the measurement data $Z$, including bus voltages, branch currents, and PQ loads,  undergoes pre-processing to transform it into a graph representation. Subsequently, the graph data is shuffled and divided into smaller batches, which are then used to train GNN. GNN is initialized and trained to find the optimal weights with back-propagation, minimizing the loss function. This iterative process continues until the desired number of training epochs is completed. The proposed PINN formulation can be described as follows
 \vspace{-4mm}
% &\hat{x}_{k} = \alpha \hat{x}_{k}^M + (1 - \alpha)\hat{x}_{k}^D, \alpha = 0.5}\\
% \begin{subequations}
% \begin{align} 
% %\label{eq:Physics-informed_Data} 
% &\hat{x}_{k}^D=F_x(\mathcal{G},W,z_k,\hat{x}_{k-1}),\\
% %\label{eq:Physics-informed_Model} 
% &\hat{x}_{k}^M= \hat{x}_{k|k-1} + K_k(z_{k} - C \hat{x}_{k|k-1}),\\
% %\label{eq:Physics-informed_Data_Fusion} 
% &\hat{x}_{k} =  \frac{\hat{x}_{k}^M + \hat{x}_{k}^D}{2},\\
% %\label{eq:Physics-informed_loss_function} 
% &\mathcal{L}(\hat{x},x)=MSE(x-\hat{x}).
% \end{align}
% \end{subequations}

\begin{subequations}
\begin{align} 
\label{eq:informed_data} 
&\hat{x}_{k}^D=F_x(\mathcal{G},W,z_k,\hat{x}_{k-1}),\\
\label{eq:informed_Model} 
&\hat{x}_{k}^O =  \frac{\hat{x}_{k}^M + \hat{x}_{k}^{D*}}{2}, \\
\label{eq:Fusion} 
& diff_{k} = \hat{x}_{k}^M - \hat{x}_{k}^{D*}, \\
\label{eq:diff} 
&\mathcal{L}(\hat{x},x)=MSE(x-\hat{x}) + MSE[diff_{k}\times(x^O-\hat{x}^O)]
\end{align}
\end{subequations}

where $x$ refers to a ground-truth state vector, $z_k$ respectively are the measurements at time step $k$. $F_x$ represents a GNN with trainable weights $W$. 

The proposed PINN formulation can be described as follows. The estimated states $\hat{x}^D$ are provided by the GNN can be described as equation (\ref{eq:informed_data}). The estimated states $\hat{x}^M$ are provided by the model-based estimation can be achieved using equation (\ref{eq:MB7}). To incorporate the model-based estimation, which contains the physics of the distribution system, with the corresponding GNN estimation $\hat{x}^{D*}$, we average these estimated states to get the overlapped states $\hat{x}^O$ and calculated the corresponding difference factors by equations \ref{eq:informed_Model} and \ref{eq:Fusion}, respectively. It should be noted that $\hat{x}^{D*}$, which shares and overlaps on state variables with $\hat{x}^M$, is a subset of GNN estimation $\hat{x}^{D}$ for entire system states.
%\vspace{-2pt}
\section{PINN-based State Estimation Implementation} 
This section presents an exhaustive overview of the implementation of physics-informed state estimation. Particularly, we discuss the selection of power systems, the appropriate algorithm, data collection, and model training strategies.
\subsection{Investigated Power Systems}
Our investigation focuses on state estimation in three-phase power systems, which are widely utilized in electrical grids. These power systems can be effectively represented as graphs due to their structures and complex component relationships \cite{liao2021review}. Graphs provide a suitable framework for modeling the power system topology, where nodes represent the buses, and edges represent the connections between them. By analyzing the dynamics and characteristics of these three-phase buses, we can gain valuable insights into power system behavior under the load-change condition \cite{tian_neural-network-based_2021}. To demonstrate the practical significance and relevance of our state estimation approach, we specifically chose the IEEE 5-bus, IEEE 123-bus, and IEEE 8500-bus systems as illustrative examples in our case studies. This choice of these systems ensures the generalizabilty and scalability of our state estimation approach, making it applicable to a wide range of power system configurations.

\RestyleAlgo{ruled}
%% This is needed if you want to add comments in
%% your algorithm with \Comment
\SetKwComment{Comment}{/* }{ */}
\begin{algorithm}[!bp] %!htp

\caption{Physics-Informed State Estimation}\label{alg:three}
\KwData{Graph $\mathcal{G}(V, E)$, measurements $z$ } 
\KwResult{States $(V, \theta)$ of all 3-phase buses}
Determine the areas for the model-based estimation;\\
Estimate $\hat{x}^M$ using the Kalman Filtering method;\\
Create a load-change scenario for different topologies;\\ 
Perform data pre-processing and normalization;\\
Define the neural network structures;\\
Initialize hyper-parameters and optimization;\\ 

\For{$epoch = 1$ \text{to} $n$}{
\For{$data$ in $dataset$}{
\If{data batch == batch size}{
Randomly sample a batch of graph data;\\
Estimate $\hat{x}^D$} (Voltage magnitude and phase) by the GNN model;\\
Fuse $\hat{x}^D$ and $\hat{x}^M$} for a strong-physics connection;\\
Calculate the loss function $\mathcal{L}(\hat{x},\: x)$; \\
Update the gradients and parameters;\\
        }
    
   Return the state variables at three-phase buses;

%\vspace{2pt}
\end{algorithm}
%
%\vspace{-5mm}
\subsection{Data Acquisition Procedure}
 The structure of a graph dataset is defined as an organized collection of a node index, node features, and label vectors. \cite{nguyen_spatial-temporal_2022}. 
 $G= \left \{ (V, x^{1}, E, y^{1}), (V, x^{2}, E, y^{2}),...,(V, x^{n}, E, y^{n}) \right \}$, where $V$ is the unchanged set of nodes, $E$ are the edges between them, and $n$ denotes the index of each graph. The node feature $x^{i} \in \mathbb{R}^{N\times F\times T}, \forall i\in \left \{ 1,...n \right \}$ contain three dimensions: the number of nodes $N$, the number of node features $F$, and the time duration $T$. The state vector of the graph network within the time duration $T$ is $y^{i} \in \mathbb{R}^{L\times P\times T}$ with $L$ as the number of state information-containing nodes and $P$ as the number of state signals. Since our current investigation only focuses on three-phase voltage measurements, we do not consider the states of single-phase buses where voltage measurements are not available. The node feature matrix encompasses the active and reactive power values throughout the power system, capturing comprehensive information across all nodes.
\vspace{1.5mm}

\begin{figure}[t!]
    \vspace{-3mm}
    \centering
    \includegraphics[height = 5.2cm, width=6.3cm]{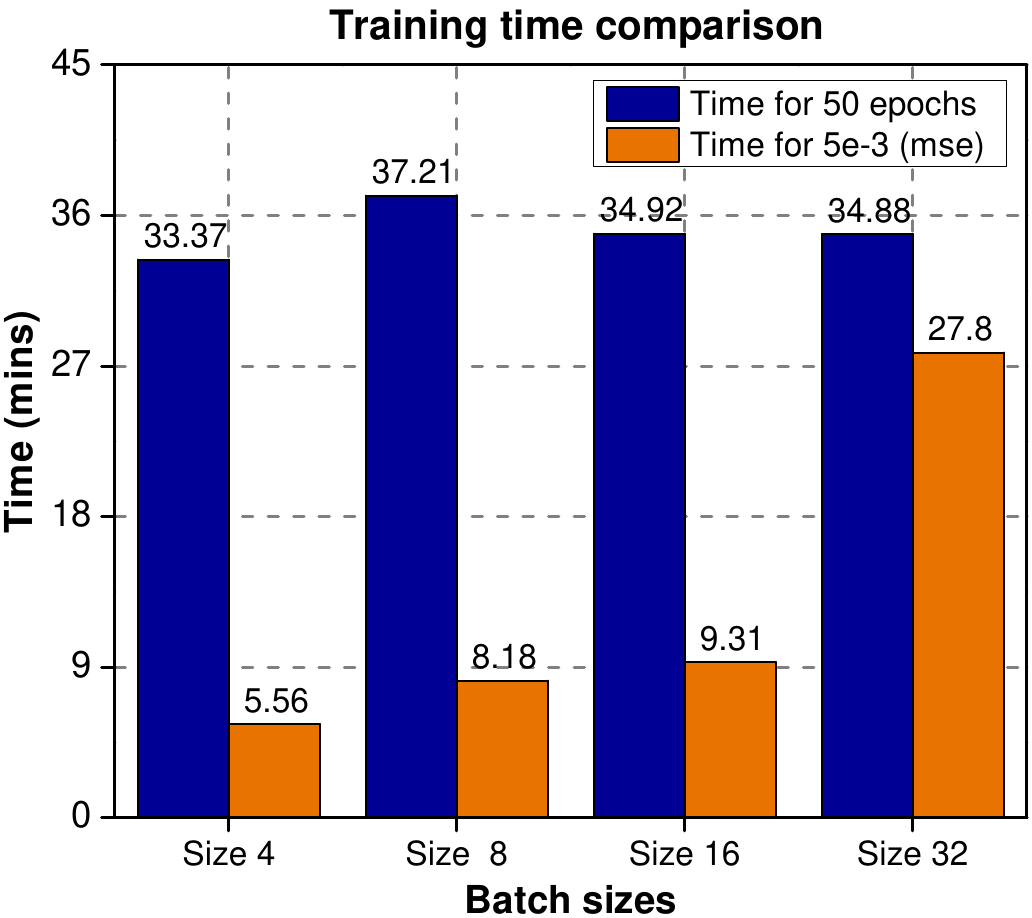}
    \vspace{-2mm}
    \caption{Training time of PINN under different batch sizes}
    \vspace{-2mm}
    \label{fig:Training_Time}
    \vspace{-3mm}
\end{figure}
\vspace{-1mm}
For the IEEE 5-bus and IEEE 123-bus datasets, measurements are gathered from the Opal-RT-Power System simulator every 600 milliseconds, with a load-change condition occurring in between. The voltage data is then transformed using Fast Fourier Transform (FFT) to obtain separate voltage magnitude and phase angle components. Alternatively, for the IEEE 8500 dataset, the data is obtained and processed through OpenDSS.  Table \ref{table:dataset} presents an overview of the case studies and datasets, categorized by system and the number of buses. The dataset allocation for each power system consists of a training set comprising 2000 graph samples, while a separate testing set is reserved, containing 500 graph samples. In addition, data inputs are normalized by the MinMaxScaler function of the Sklearn library. A Python script with RT-LAB API is written and run with RT-LAB to automatically collect the operational data of the IEEE 5-bus and the IEEE 123-bus and store them in specific folders. The data collected in the MAT-file format from Opal-RT comprises instantaneous voltage and current values of the main nodes\cite{nguyen_spatial-temporal_2022}. The Python scripts are written to process the raw data into a ready-to-use format compatible with open-source frameworks, ensuring convenience in both usage and storage.

% \vspace{1mm}
In terms of the model-based state estimation, we conduct the example on the IEEE 5-bus system, considering the branch current model between bus 3 and bus 4. The measurements are added white Gaussian noises of 5e-4 p.u. These values are chosen based on the precision (±0.05\% true value error of PMU). To assess the estimation accuracy, the means squared error (MSE) defined as

 \vspace{-1mm}
\begin{equation} MSE_{k} = \frac{1}{m}(\hat{v_{k}} - v_{k})^{T}(\hat{v_{k}} - v_{k})
\end{equation}

where $k$ represents the time step, $m$ is the number of signals, $\hat{v}_{k}$ is the estimated bus voltages, and $v_{k}$ is the true states of bus voltages.

The state estimation results achieved using PINN are compared with other machine learning networks, such as MLP and GNN. Table III provides an overview of the neural network architectures for state estimation in the IEEE 123-node system, highlighting the functional layers and tensor sizes. The input comprises load values from 51 three-phase buses. To avoid negatively affecting the estimation accuracy, the missing node features at the buses without loads are removed. As a result, 124 input features are obtained for the training process. Reshaping and flattening operations are utilized to ensure compatibility between layers. Importantly, since PINN incorporates graph learning and model-based estimation, the number of hidden nodes in each layer of PINN exhibits the same structure as the ones of the stand alone GNN.

\subsection{Response to Load Changes}
In a real-time power system operation, the system behavior is always changing due to the variation of load. Therefore, it is necessary to consider load changes for the real-time state estimation. To ensure the effectiveness of state estimation methods, it is imperative that they can accurately capture the rapid dynamics associated with state changes \cite{tian_neural-network-based_2021}. In load-change scenarios, the active and reactive power values are intentionally changed randomly, ranging from 70\% to 130\% of the default loads. Such intentional load changes cause transient fluctuations in the actual states, further challenging the state estimation process. By designing and conducting various load-change conditions, the robustness of state estimation methods can be thoroughly evaluated, providing valuable insights into the ability of the algorithms to effectively capture the transient state fluctuations in dynamic power systems.
\vspace{-2mm}

\renewcommand{\arraystretch}{1.25}
\begin{table}[!bp] %!htp
\vspace{-2mm}
\centering % instead of \begin{center}
    \caption{CASE STUDIES AND DATASETS OVERVIEW}
    \addtolength{\tabcolsep}{-2pt}
\begin{tabular}{ *5c } %|c|c|c|c|
\hline
\textbf{ Systems} & \textbf{ Buses} & \textbf{Count} \\
\hline
\multirow{3}{1 em}

IEEE 5-bus & \:\:\: 2, 3, 4, 5    & 4  \\

\hline
\multirow{3}{1 em}{\textbf{}} & \:\:\:\:\: 1, 7, 8, 13, 18, 21, 23, &  \\ 
&  \:\:\:\:\:\:25, 28, 29,  30, 35, 40, & 51  \\ 
\:\:\:IEEE 123-bus & \:\:\:\:\: 42, 44, 47, 49, 50, 51, &  \\ 
& \:\:\:\:\:   52, 53, 54, 55, 57, 60, & \\
& \:\:\:\:\:  62, 63, 64, 65, 67, 72, &\\
& \:\:\:\:\: 76, 77, 78, 80, 81, 82, &\\
& \:\:\:\:\:  83,  86, 87, 89, 91, 93, &\\
& \:\:\:\:\: 97, 98,  99, 100, 101, &\\
& \:\:\:\:\: 105, 108, 197, &\\
\hline

\multirow{3}{1 em} \textbf{IEEE 8500-bus} & \:\:\:\:\: All 3-phase buses & 648 \\ 
\hline

\multicolumn{3}{c}{\textbf{IEEE 5-bus LC cases}: 2500 samples $|$ \textbf{Train}: 2000 $|$
\textbf{Test}: 500} & \\
\hline

\multicolumn{3}{c}{\textbf{IEEE 123-bus LC cases}: 2500 samples $|$ \textbf{Train}: 2000 $|$
\textbf{Test}: 500} & \\
\hline

\multicolumn{3}{c}{\textbf{IEEE 8500-bus LC cases}: 2500 samples $|$ \textbf{Train}: 2000 $|$
\textbf{Test}: 500} & \\

\hline
\end{tabular}
%\vspace{-2mm}
\label{table:dataset}
\end{table}

\vspace{-2mm}
\begin{figure}[t!]
    %\vspace{0.5mm}
    \centering
    \includegraphics[height = 4.9cm, width=6.6cm]{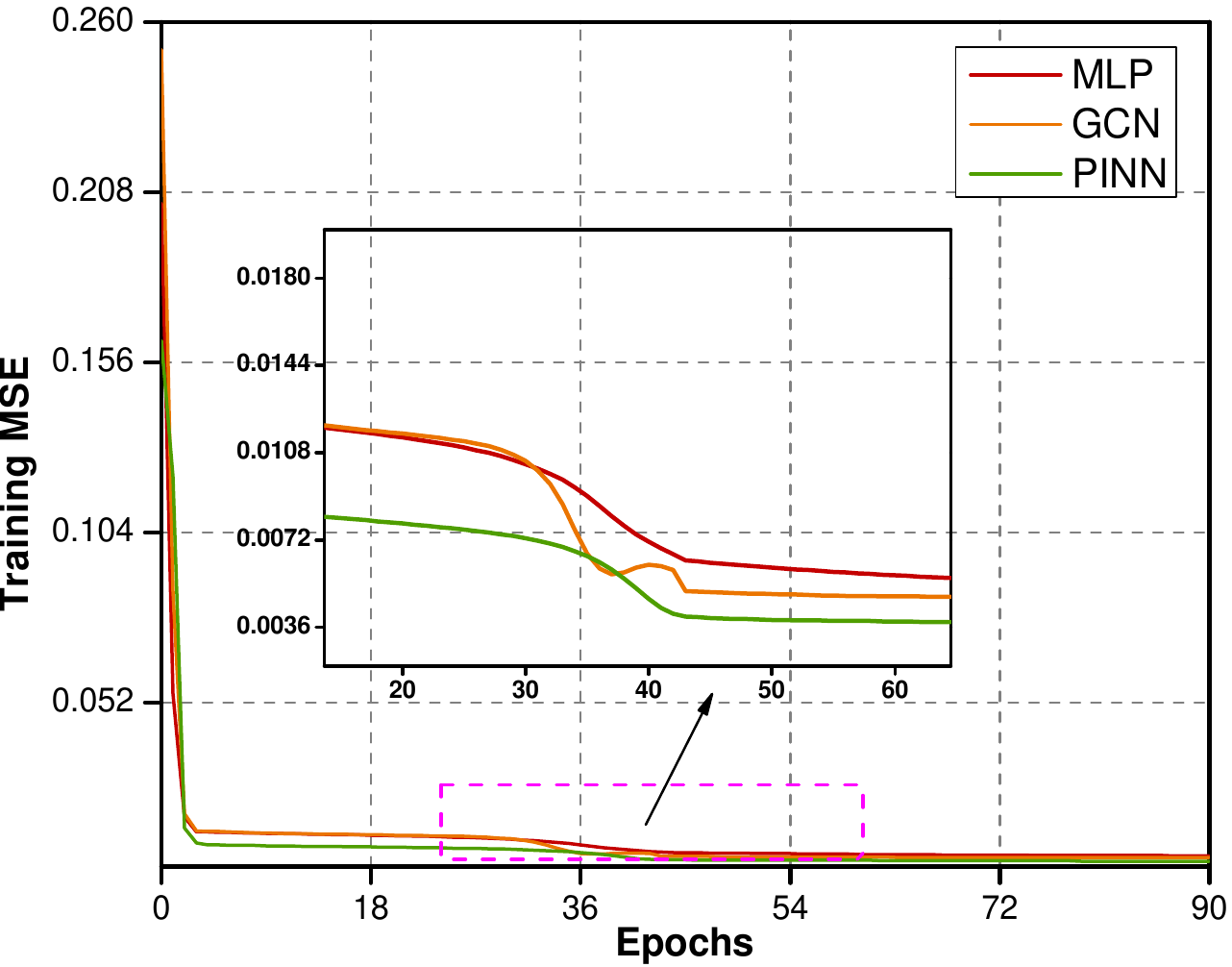}
    \vspace{-4mm}
    \caption{Training MSE curve in the IEEE 123-node system}
    \vspace{-3mm}
    \label{fig:Training_MSE}
    %\vspace{-0.5mm}
\vspace{-1mm}
\end{figure}
\subsection{Configuration of Training Model}
The flowchart, shown in Fig. \ref{fig:flow_chart},  provides a detailed the physics-informed state estimation process in power systems, including both offline and online training stages. During offline training, the model parameters are optimized using historical data to capture the system's underlying dynamics and relationships. After training models in the offline phase, these models can be applied to estimate states under different topologies using real-time measurements obtained from monitoring devices such as PMUs or SCADA, enhancing accuracy and adaptability. This approach ensures that the method is well-informed by historical data while being capable of performing state estimation in real-time scenarios.

\vspace{1.5mm}
%\subsection{Configuration of Training Model}
%\vspace{3mm}
\begin{figure}[t!]
    %\vspace{-3mm}
    \centering
    \includegraphics[height = 5.5cm, width=7.0cm]{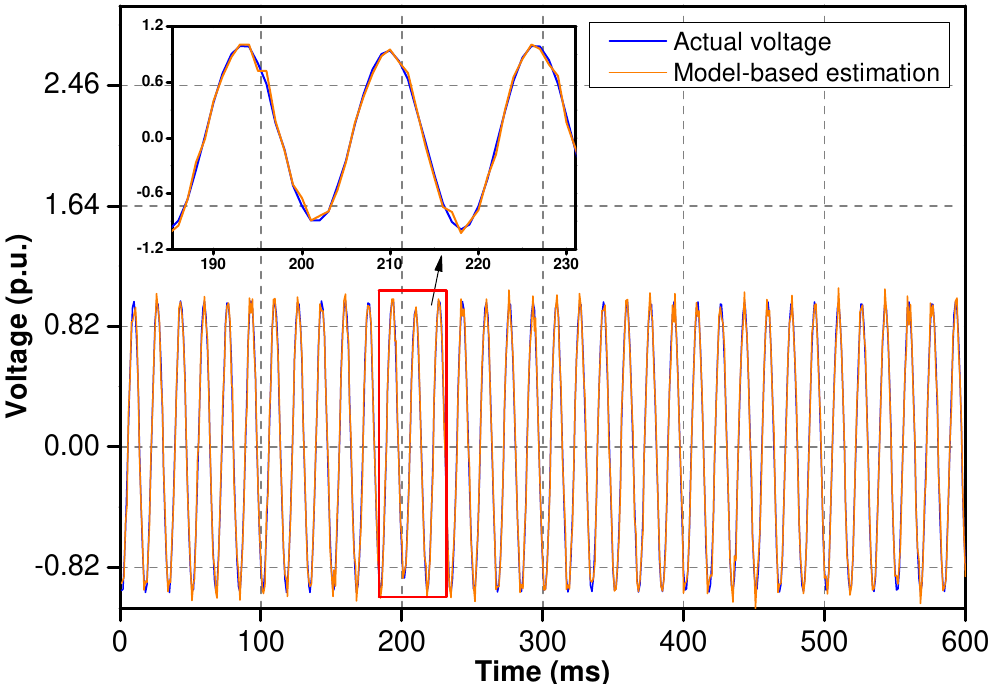}
    \vspace{-2mm}
    \caption{Model-based estimation of $v_{4a}$ in IEEE 5-bus system}
    %\vspace{-2mm}
    \label{fig:Model_based_v4a}
    \vspace{-1mm}
%\vspace{-3mm}
\end{figure}
\begin{figure}[t!]
    %\vspace{-3mm}
    \centering
    \includegraphics[height = 5.5cm, width=7.0cm]{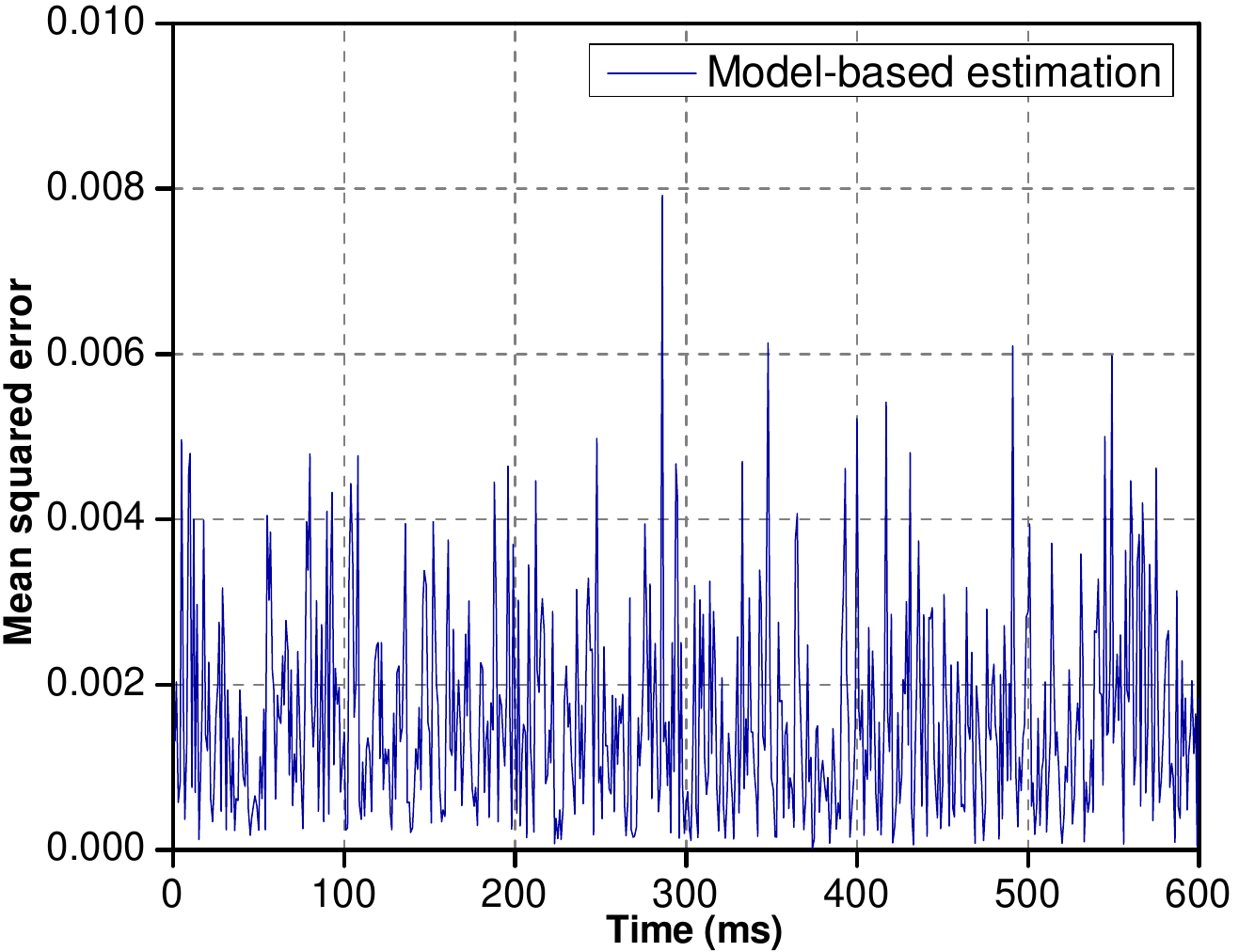}
    \vspace{-2mm}
    \caption{Mean squared errors of model-based estimation}
    %\vspace{-2mm}
    \label{fig:Model_based_mse}
    \vspace{-2mm}
\vspace{-3mm}
\end{figure}
The graph datasets are trained with the Stochastic Gradient Descent (SGD) optimizer using mean square error as the estimation metric. This choice is common for regression tasks and signifies a desire to minimize the squared differences between predicted and actual values. To address the over-fitting problem, the dropout layers are included in hidden layers\cite{labach2019survey}. The batch size serves as a crucial parameter that significantly impacts both the execution time and model accuracy. The large batch size leads to a precise estimation of the gradient but the time consumption increases remarkably\cite{masters2018revisiting}. For graph datasets, the batch size should be chosen carefully, considering the number of nodes in each graph to ensure efficient training. The other hyper-parameters such as model architecture, learning rate, and model complexity are selected based on our experience. The learning rate is within a range of 0.05 to 0.001 in conjunction with the SGD optimizer. The training process is executed on a computer equipped with an AMD EPYC 7302 16-Core Processor (3.0 GHz and 256 GB RAM). The deep learning framework utilized for the study is PyTorch, specifically leveraging the PyTorch Geometric library, which is tailored for GNN model tasks\cite{fey2019fast}.

Fig. \ref{fig:Training_Time} illustrates the comparison of time consumption trained by PINN in the IEEE 5-bus system to complete 50 epochs and achieve a specific loss value as 5e-3 across different batch sizes. Overall, it can be seen that the duration to complete both tasks varies depending on the chosen batch size. For the first 50 epochs, a batch size of 4 takes approximately 33.37 minutes. Meanwhile, batch sizes of 8, 16, and 32 require 37.21 minutes, 34.92 minutes, and 34.88 minutes respectively. In terms of achieving a sufficiently low loss, a batch size of 4 takes around 5.56 minutes. Batch sizes of 8, 16, and 32 require 8.18 minutes, 9.31 minutes, and 27.8 minutes respectively. Considering both aspects, a batch size of 4 appears to be optimal for the training. However, the selection of a batch size of 4 is relative and requires further investigation to determine the optimal parameter for enhancing the model's performance. Similar considerations were made for batch size selection in the training process of the two remaining systems.

\begin{figure}[t!] %htp
    %\vspace{-3mm}
    \centering
    \includegraphics[height = 5.3cm, width=7.0cm]{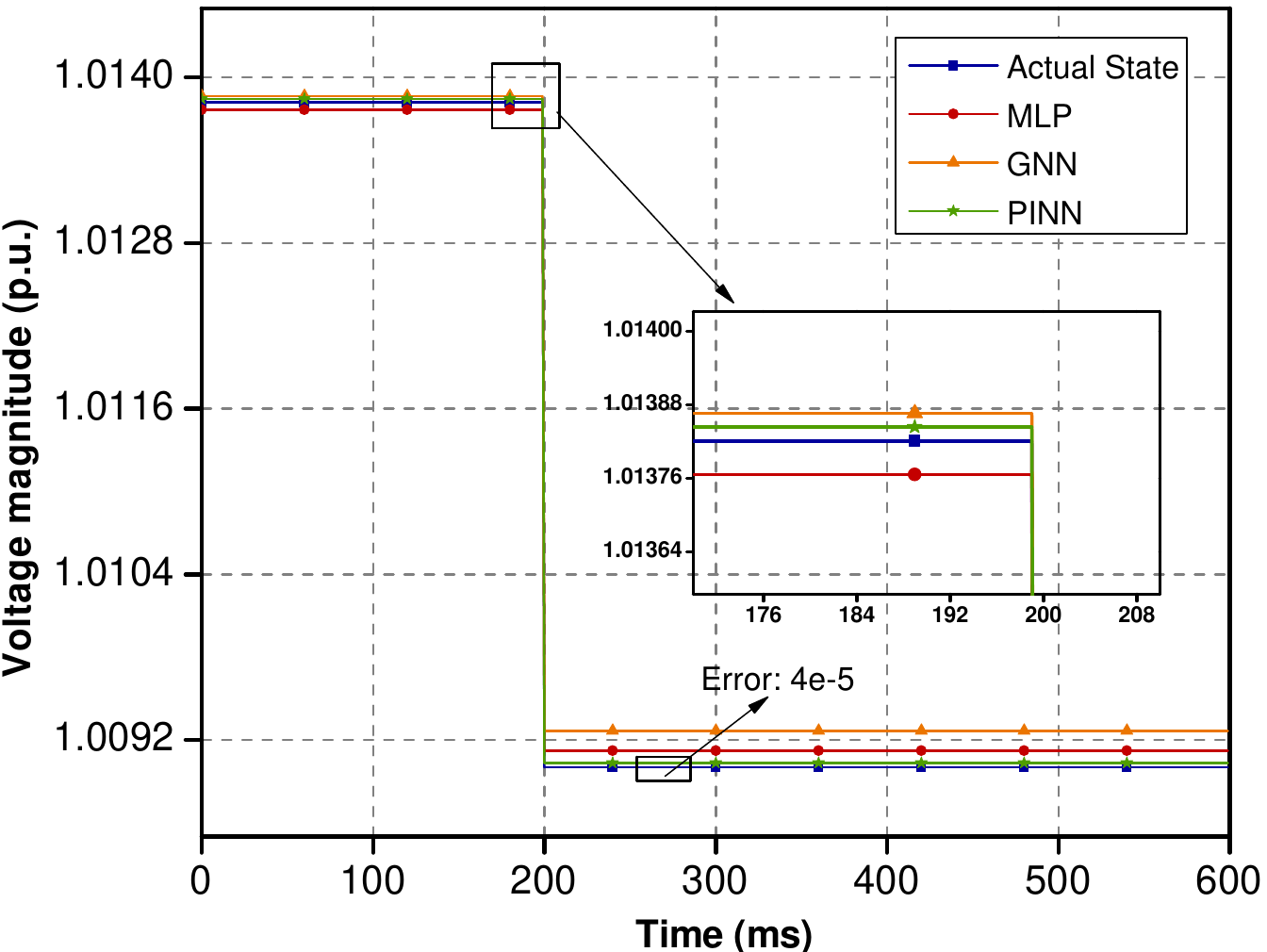}
    \vspace{-2mm}
    \caption{Voltage magnitude estimation on IEEE 5-bus system}
    %\vspace{-2mm}
    \label{fig:IEEE5vSE}
    %\vspace{-1mm}
\end{figure}
\begin{figure}[t!]
    %\vspace{-3mm}
    \centering
    \includegraphics[height = 5.5cm, width=7.0cm]{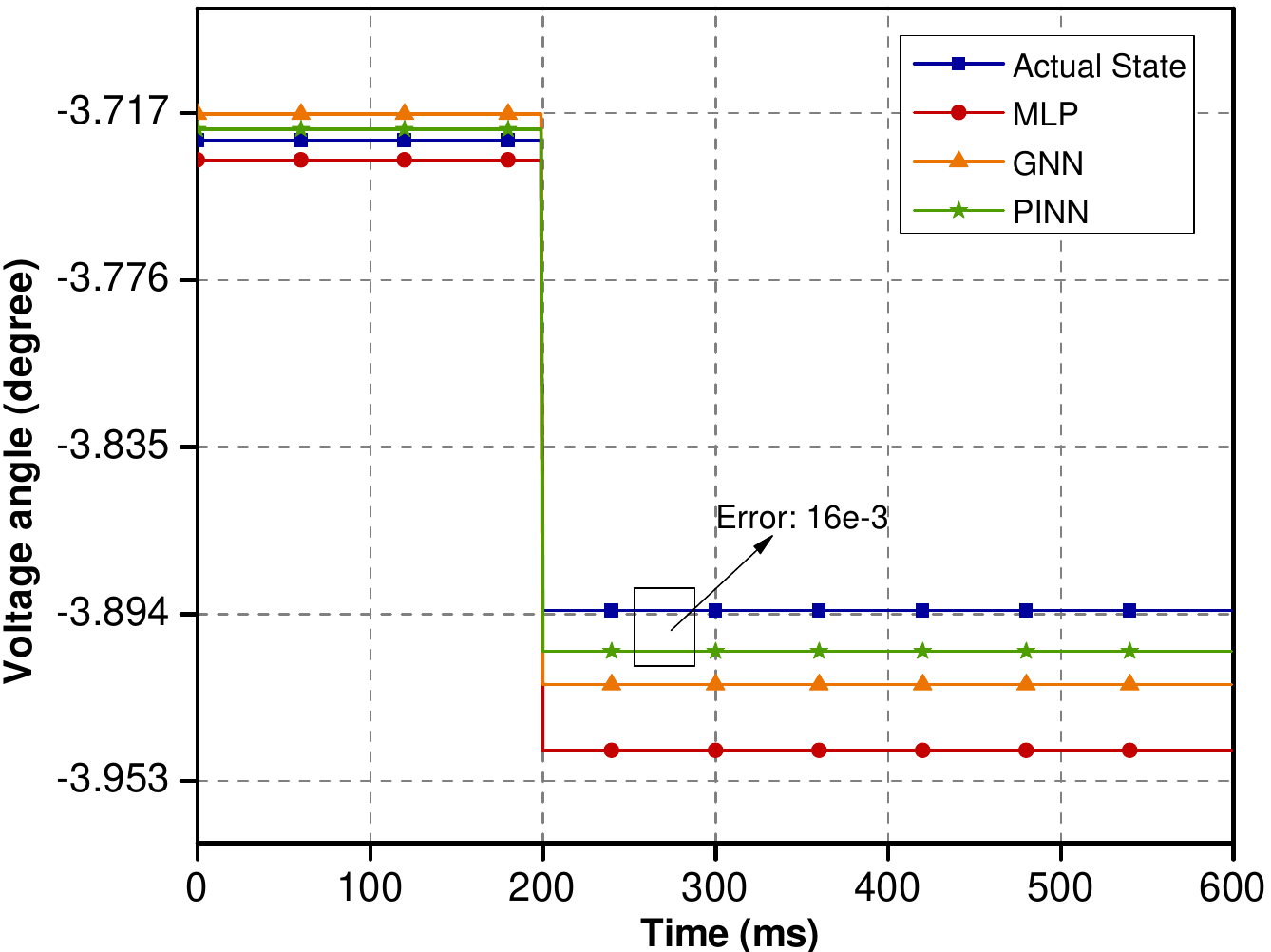}
    \vspace{-2mm}
    \caption{Voltage phase estimation on IEEE 5-bus system}
    %\vspace{-2mm}
    \label{fig:IEEE5pSE}
    \vspace{-5mm}
\end{figure}

\section{RESULTS AND ANALYSIS} 
This section provides a comprehensive analysis of the results obtained from applying the proposed method on three different power systems: IEEE 5-bus, IEEE-123 bus, and IEEE 8500-bus. Furthermore, we include results of the comparative study, presented in Table \ref{table:loss}, to evaluate the performance of the proposed method and alternative machine learning techniques in state estimation.
\vspace{-1mm}
\begin{figure}[!t]
    \vspace{-5mm}
    \centering
    \includegraphics[height = 3.6cm, width=5.6cm]{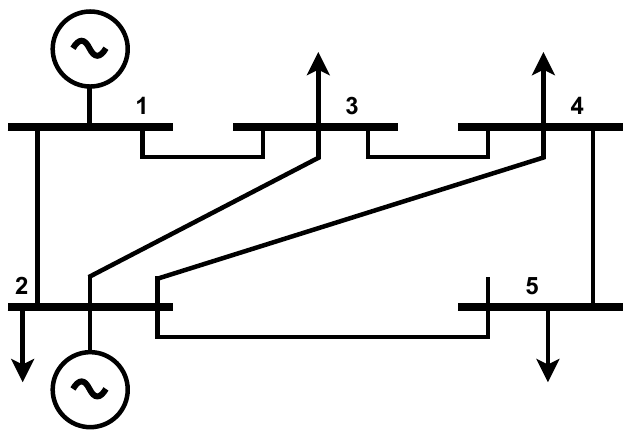}
    \vspace{-2mm}
    \caption{The topology of IEEE 5-bus network}
    %\vspace{-2mm}
    \label{fig:IEEE_5bus}
    %\vspace{-2mm}
\end{figure}
\begin{figure}[!t]
    %\vspace{-3mm}
    \centering
    \includegraphics[height = 5.3cm, width=6.0cm]{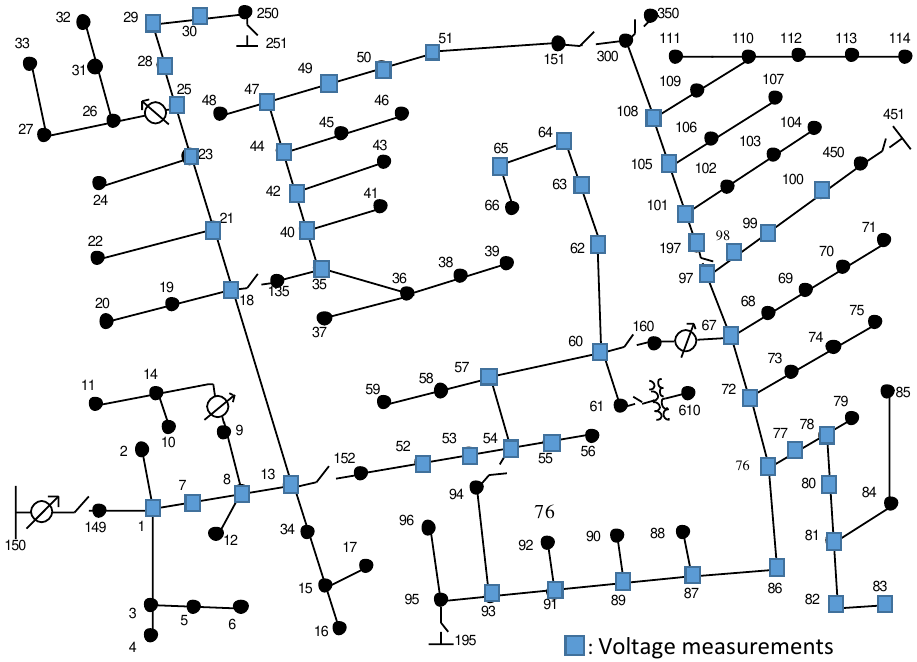}
    \vspace{-2mm}
    \caption{The topology of IEEE 123-bus distribution network}
    %\vspace{-2mm}
    \label{fig:IEEE_123bus}
    \vspace{-4mm}
\end{figure}
\subsection{Model-based Method Accuracy}
Fig. 7 displays an estimation of a sinusoidal voltage $v_{4a}$ under a load-change scenario through the Kalman filter method which exhibits a notable proximity to the true values, with the orange line representing the estimated values and the blue line depicting the actual values. The resulting MSE values are presented in Fig. \ref{fig:Model_based_mse}. Notably, the MSE values for estimated and actual states during 600 ms remain below 0.008, even under the condition of a load-change scenario. Comparatively, the model-based estimation conducted on the Potsdam 13-bus system in \cite{nguyen2021distributed} obtained MSE values of less than 0.01. Therefore, the result provides robust verification of our model-based estimation algorithm.

\vspace{-2mm}
\renewcommand{\arraystretch}{1.1}
\begin{table}[!bp] %[!htp]  table*} [t]
\centering
\caption{Comparison of neural network structures}
\addtolength{\tabcolsep}{-0.1pt}
  \begin{tabular}{|c|c|c|c|c|c|c|} %|c|c|c|c|c|c|c|
    \hline
    \multicolumn{6}{|c|}{\textbf{Our implementation}} \\
    \hline
    % \multirow{2}{*}{Dataset} &
      \multicolumn{2}{|c|}{MLP} &
      \multicolumn{2}{c|}{GNN} &
      \multicolumn{2}{c|}{PINN}\\
    \hline
    Input & [600$\times$124] & Input & [600$\times$124] & Input & [600$\times$124] \\
    \hline
   
     Dense & [1024] & Dense & [1024] & Dense & [1024] \\
    \hline
    Dense & [1024] & Dense & [1024] & Dense & [1024] \\
    \hline
    % \multicolumn{2}{|c|}{Params: xxxx} &
    %   \multicolumn{2}{c|}{Params: xxxx} &
    %   \multicolumn{2}{c|}{Params: xxxx}\\
    %   \hline
    Dense & [640] & GNN & [1024$\times$640] & GNN & [1024$\times$640] \\
    \hline
    Dense & [306] & GNN & [640$\times$306] & GNN & [640$\times$306] \\
    \hline
    Dense & [306] & Dense & [306] & Dense & [306] \\
    \hline

    \multicolumn{6}{|c|}{\textbf{Reference implementation}} \\
    \hline
      \multicolumn{2}{|c|}{SNN \cite{8962079}} &
      \multicolumn{2}{c|}{LSTM \cite{8962079}} &
      \multicolumn{2}{c|}{MLP \cite{azimian_state_2022}}\\
    \hline
    Input & [190] & Input & [190] & Input & [113] \\
    \hline
   
     Dense & [256] & Dense & [256]  & Dense & [800] \\
    \hline
    Dense & [920] & LSTM & Not stated &  Dense & [800] \\
    \hline

    Dense & [920] & LSTM & Not stated & Dense & [800] \\
    \hline
    Dense & [920] & LSTM & Not stated & Dense & [800] \\
    \hline
    Dense & [494] & Dense & [494] & Dense & [924] \\
    \hline
    
  \end{tabular}
  \label{table:structures}
  \vspace{-1mm}
\end{table}

\vspace{1mm}
\subsection{IEEE 5 Buses}
Fig. \ref{fig:IEEE_5bus} displays the IEEE 5-bus system comprising 2 generators located at bus 1 and bus 2, where bus 1 functions as a slack bus and bus 2 is considered as a bus with negative load. The input features include injection P and Q at each bus, while the outputs consist of system states at load buses. Fig. 9 and 10 compare the actual values (in blue lines) with the voltage magnitude and phase angle estimations produced by PINN (in green lines), MLP (in red lines), and GNN (in orange lines) at bus 2 when a load-change scenario occurs at the 200ms time mark. As shown in these two figures, the estimations of PINN, including the voltage magnitude that decreases from approximately 1.0138 p.u to 1.0092 p.u and the phase angle that shifts from -3.725 to -3.895, are closer to the true values than both the estimations provided by MLP and GNN. The estimation results on the IEEE 5-bus dataset are outlined in Table \ref{table:loss}.

\begin{figure}[t!]
    \vspace{-3mm}
    \centering
    \includegraphics[height = 5.8cm, width=5.7cm]{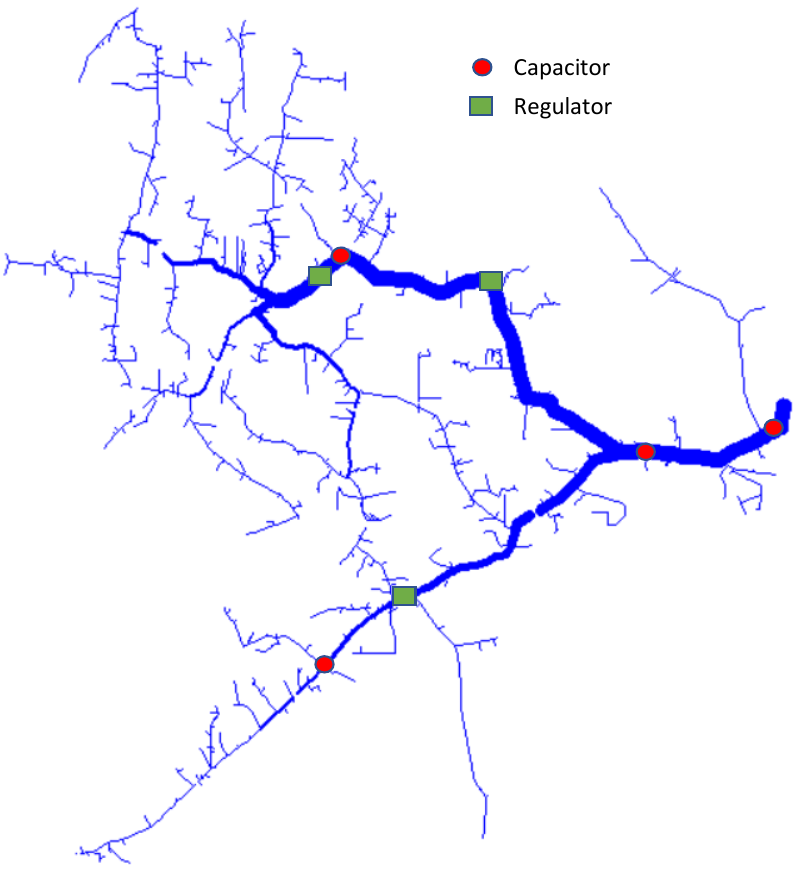}
    \vspace{-2mm}
    \caption{The topology of IEEE 8500-bus network}
    %\vspace{-2mm}
    \label{fig:IEEE_8500bus}
    \vspace{-3mm}
\end{figure}
\subsection{IEEE 123 Buses}
%%
%\vspace{-1mm}
The algorithm's validation is extended to the medium-size system of the IEEE 123-node distribution test feeder, shown in Fig. \ref{fig:IEEE_123bus}. In this distribution system, voltage measurements are collected at buses marked with a blue square, and we intentionally reduce the number of voltage measurement inputs to assess the trained state estimation models' performance. The training curve of the IEEE 123-node system is depicted in Fig. \ref{fig:Training_MSE}. Remarkably, all three models' training curves converge after 150 episodes, with PINN achieving the lowest mean squared error (MSE) loss of 0.0032 compared to the other models. The slight drop in the training curve at epoch 40, attributed to a decrease in the learning rate from 0.05 to 0.01, is a planned adjustment to fine-tune training. In Fig. 14 and 15, voltage magnitude and phase angle estimations at bus 21 changing from 0.9762 p.u to 0.9782 p.u and approximately -1.88 to -1.71, respectively, show that PINN provides closer approximations to actual states than MLP and GCN on the IEEE 123-bus dataset.
%\vspace{1mm}
\subsection{IEEE 8500 Buses}
The final case study involves the IEEE 8500-node distribution system, as depicted in Fig. \ref{fig:IEEE_8500bus}. Due to the substantial number of output channels, we conducted separate state estimations for voltage magnitude and phase angle in this scenario. Fig. 16 and 17 showcase the performance of three models: MLP, GNN, and PINN, concerning bus $m1047763$ in the IEEE 8500-node system. The voltage magnitude, estimated by PINN, at this bus rises from 0.9961 p.u to 0.9996 p.u, with a corresponding phase shift from approximately -38.61 to -38.57, outperforming both GNN and MLP. After 150 epochs, PINN emerges as the top performer, achieving MSE value of 0.0039 for voltage magnitude and 0.0005 for phase angle, as summarized in Table \ref{table:loss}.

\begin{figure}[!t]
    \vspace{-1.0mm}
    \centering
    \includegraphics[height = 4.8 cm, width=7.0cm]{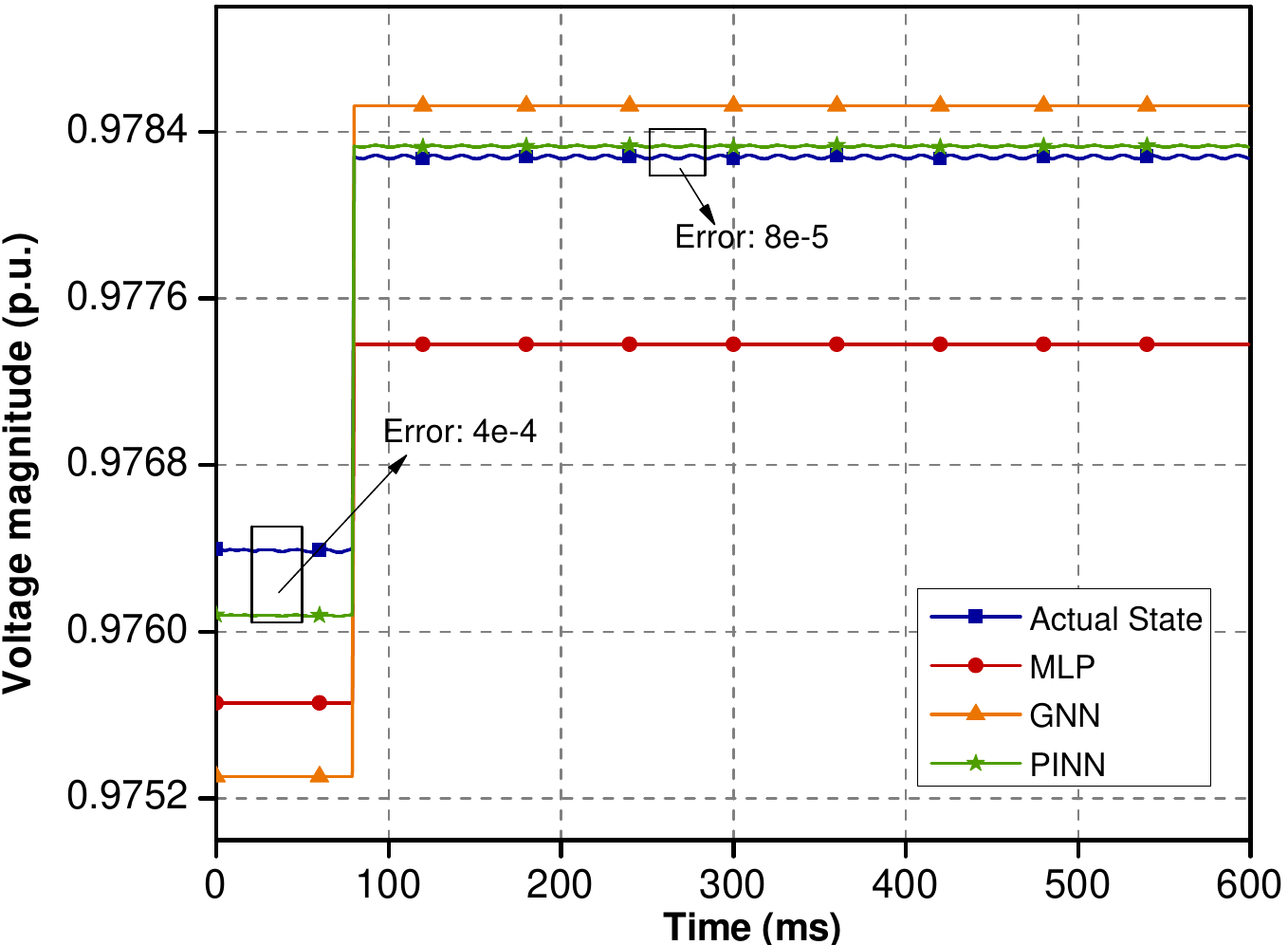}
    \vspace{-2mm}
    \caption{Voltage magnitude estimation on IEEE 123-bus system}
    %\vspace{-2mm}
    \label{fig:IEEE123vSE}
    \vspace{-1mm}

\end{figure}
\vspace{2mm}
\begin{figure}[!t]
    \vspace{-1mm}
    \centering
    \includegraphics[height = 5.0cm, width=7.0cm]{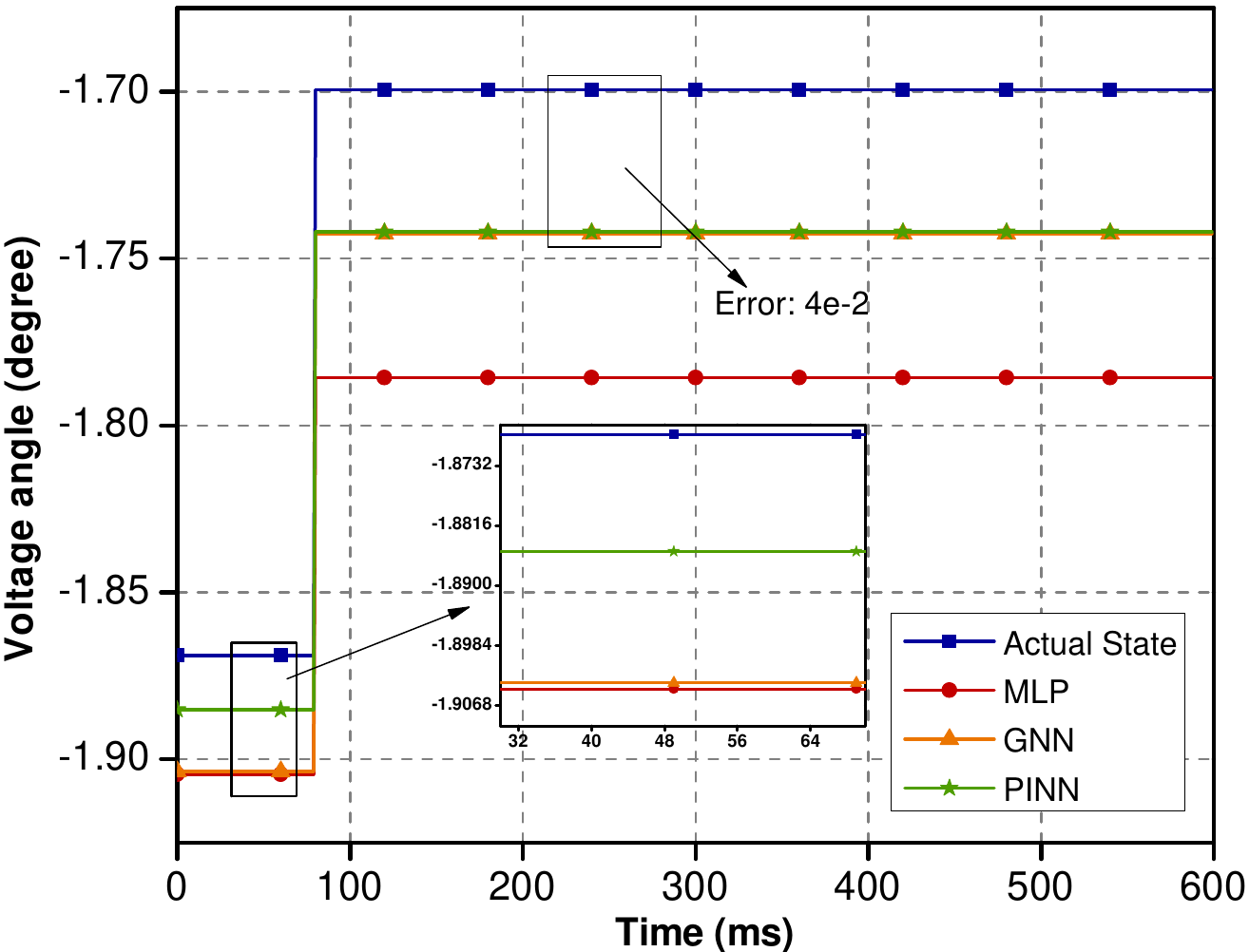}
    \vspace{-2mm}
    \caption{Voltage phase estimation on IEEE 123-bus system}
    %\vspace{-2mm}
    \label{fig:IEEE123pSE}
    \vspace{-3mm}

\end{figure}
\vspace{-3mm}
\subsection{Discussion and Limitations}
\begin{enumerate}[leftmargin=0cm,itemindent=0.9cm,labelwidth=\itemindent,labelsep=-0.4cm,align=left]

\item \textit{Architecture complexity}: Looking at Table III, the proposed PINN implementation for the 123-bus system utilizes 5 hidden layers, each containing 306, 640, or 1024 nodes. Considering state-of-the-art methods, \cite{8962079} employs an LSTM with 3 layers, trained with 7 auto-encoders and a shallow neural network with 920 neurons in the hidden layers to validate state estimation in IEEE 123-bus system. Similarly, \cite{azimian_state_2022} focuses on a case study of the 240-Node Network in the Midwest U.S., estimating 924 states with an MLP of 5 layers, each containing 800 hidden nodes. These data-driven models utilize a significant number of nodes within their hidden layers. The underlying reason is that the inputs of the training models are derived from measurement data, while the outputs correspond to the total number of states that need to be estimated. As power systems scale up, both the number of layers and hidden nodes tend to increase, resulting in increased model complexity. Assuming $n$ represents the number of power system buses, the computational complexity of the proposed method is nearly quadratic, $O(n^{2})$. Thus, it is challenging to apply physics-informed neural network to real-world scenarios with millions of nodes. Implementing federated learning is a potential direction to distribute and scale up training. 
\begin{figure}[!t]
    %\vspace{-3mm}
    \centering
    \includegraphics[height = 4.8cm, width=7.0cm]{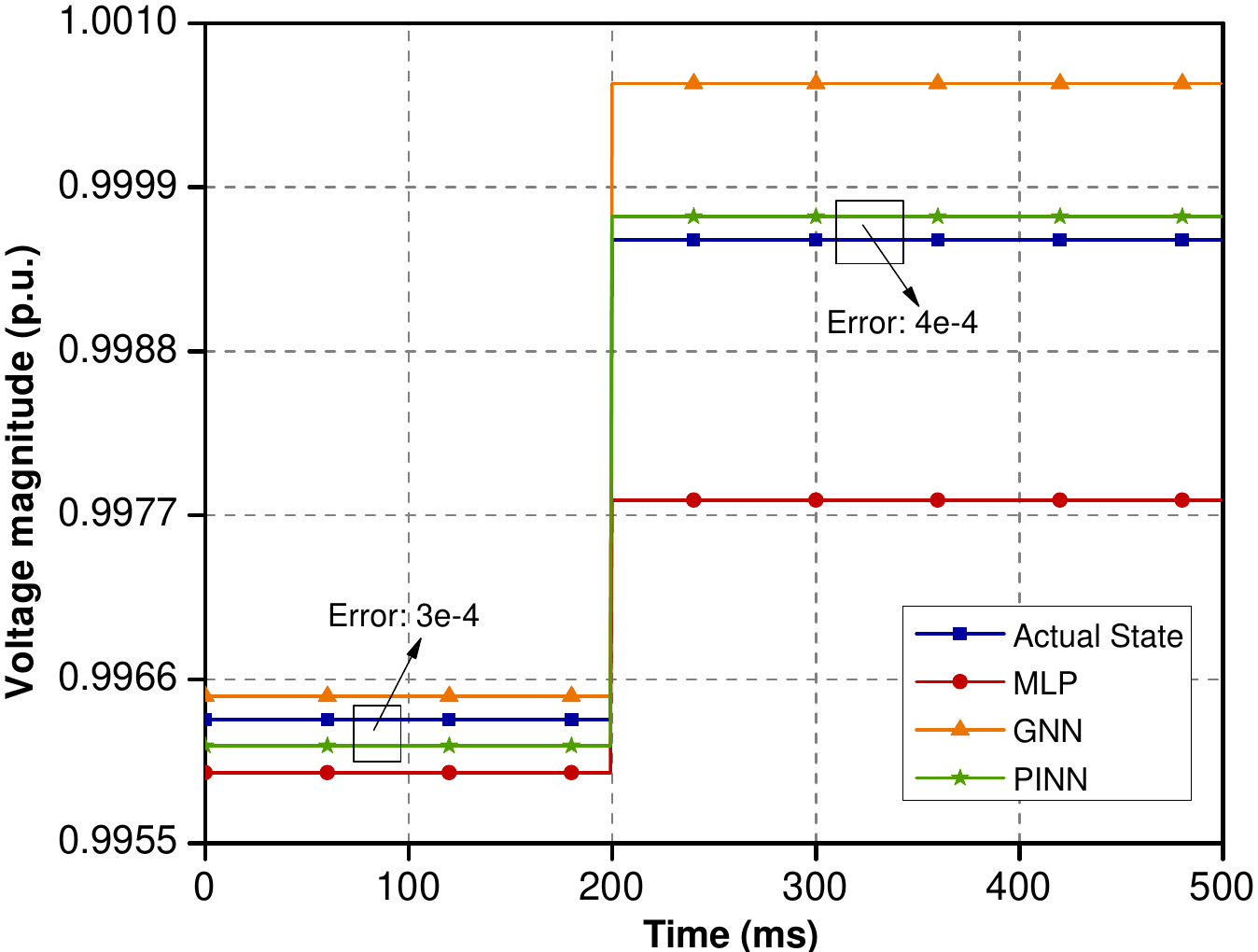}
    \vspace{-2mm}
    \caption{Voltage magnitude estimation on IEEE 8500-bus system}
    %\vspace{-2mm}
    \label{fig:IEEE8500VSE}
    \vspace{-3mm}
\end{figure}
\begin{figure}[!t]
    %\vspace{-0.7mm}
    \centering
    \includegraphics[height = 4.9cm, width=7.0cm]{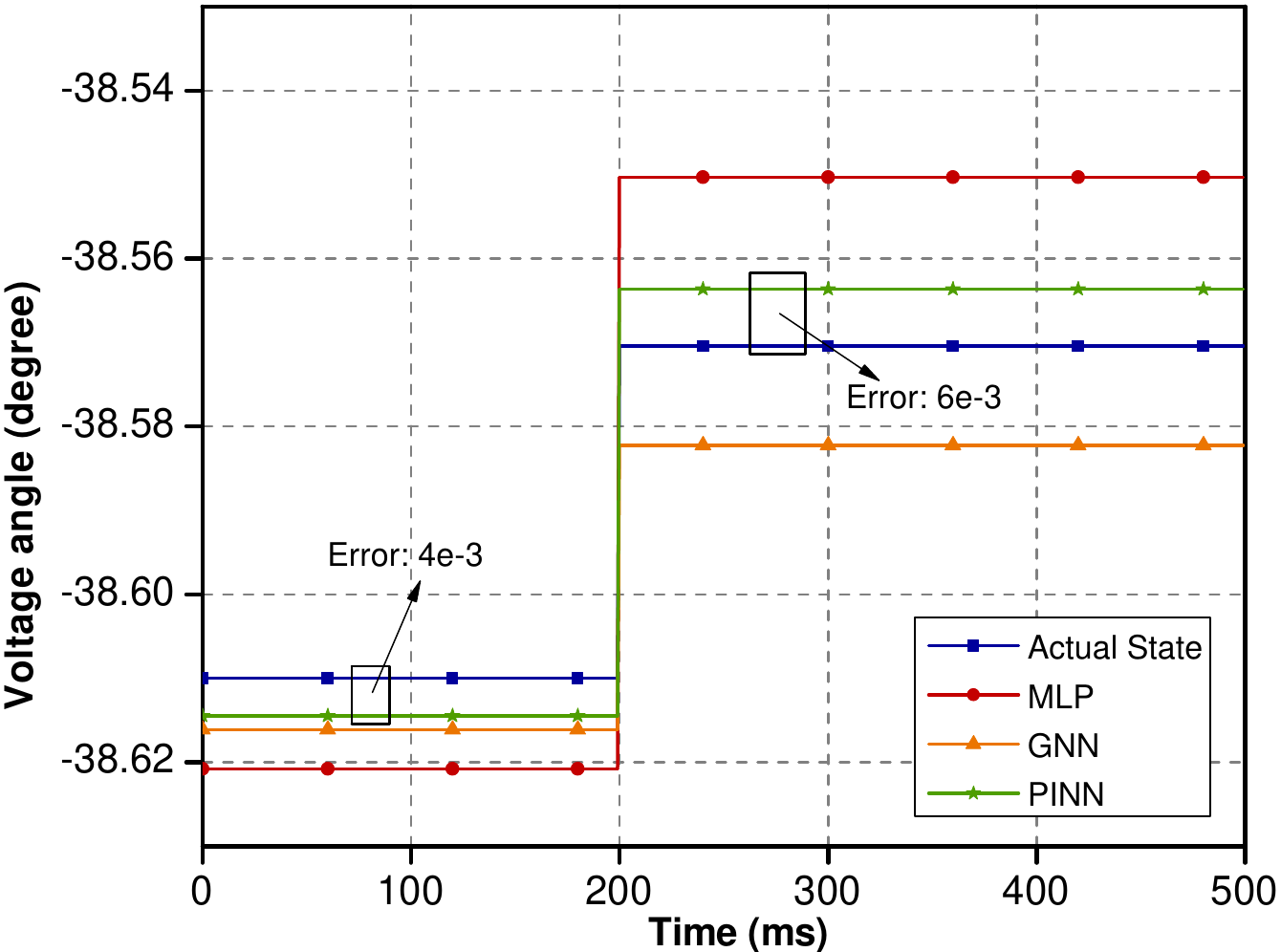}
    \vspace{-2mm}
    \caption{Voltage phase estimation on IEEE 8500-bus system}
    %\vspace{-2mm}
    \label{fig:IEEE8500pSE}
    \vspace{-3mm}
\end{figure}
\vspace{1mm}
\setlength{\tabcolsep}{1pt}
\renewcommand{\arraystretch}{1.3}
\begin{table}[!bp]
\caption{Performance comparison of different approaches}
\vspace{-5mm}
\label{T:equipos}
\begin{center}
\begin{tabular}{| c | c | c | c| c|}
\hline
\textbf{ Model } & \multicolumn{4}{ c |}{\textbf{Criteria (Training 1/ Training 2) }}  \\ 
\cline{2-5}
& \textbf{IEEE 5-bus} & \textbf{ IEEE 123-bus\:} & \textbf{ 8500-bus Mag\:} & \textbf{ 8500-bus Phase\:}\\
\hline

MLP & \:3.0e-4/ 2.0e-4\:& 0.0051/ 0.0049 & 0.0161/ 0.0178  & 0.0011/ 0.0009\\ \hline

GNN & 2.0e-4/ 1.0e-4 & 0.0040/ 0.0045 & 0.0172/ 0.0109& 0.0008/ 0.0010\\ \hline

PINN & \textbf{8.7e-5}/ 9.6e-5& 0.0034/ \textbf{0.0032}& 0.0055/ \textbf{0.0039} & 0.0007/ \textbf{0.0005}  \\ \hline
\end{tabular}
\end{center}
\vspace{-3mm}
\label{table:loss}

\end{table}
%%%%
\item \textit{Performance analysis}: Looking at Table \ref{table:loss}, our proposed method outperforms alternative state estimation techniques, achieving a lower validation mean square error. In the IEEE 5-bus transmission system tests, the PINN achieved remarkably low testing errors, reaching as low as 8.7e-5, thereby surpassing the performance of both MLP and GNN. In the case of the IEEE 123-bus system, the PINN achieved a 28$\%$ lower MSE than the GNN, while on the IEEE 8500-bus system, it outperformed the GNN by achieving a 60$\%$ reduction in MSE.  These validation results on three IEEE test systems not only indicate the adaptability of PINN in different operational environments but also highlights its effectiveness in enhancing state estimation. However, both grid topologies, the sizes of the neural networks, and training parameters influence the validation metric. For the balanced-load transmission system like the IEEE 5-bus, we consider only a single phase as representative, simplifying the approach but in the case of the two distribution systems, all three phases should be estimated, leading the difference MSE results on systems. Additionally, looking at Fig. \ref{fig:Training_MSE}, the initial slope in the training curve's early epochs indicates rapid learning of the models, though it may also suggest potential issues like suboptimal initialization or an excessively high learning rate. Therefore, in future research, the training process could be further improved selection of optimal hyper-parameters.
%\vspace{-1mm}

\item \textit{Limitations}: Although our results have demonstrated the PINN performance, there are some limitations that remain to be addressed. (1) Estimating an expansive number of states poses challenges in employing our approach to large systems with millions of nodes. (2) State estimation under noise measurements or disturbances from cyber-attacks have not been considered for real-world operation. (3) The weighting hyper-parameters of the training process and the fusion formulation between terms should be optimized in future works.
\end{enumerate}

\section{Conclusions} 
This paper presents a physics-informed graph neural network-based approach to enhance power system state estimation with limited measurements. The main contribution is the novel integration of a graph neural network with a model-based Kalman filter to perform dynamic state estimation with a limited dataset of only 2500 samples, which can be prepared in practice. The Kalman filter produces the physical knowledge through the branch current formulation, and GNN captures spatial-temporal patterns from graph-structured data. The proposed method estimates voltage magnitudes and phase angles from the measurements by minimizing a physics-informed loss function, updating the weights of underlying neural networks, and leveraging the messaging-passing mechanism of GNN. We experimentally demonstrated that our proposed physics-informed graphical learning method achieves lower MSE more than 20$\%$ compared to existing alternatives such as MLP and GNN on three test standard IEEE grids. In our future works, the PINN with distributed training under attacks and noises scenarios will be developed.

% \section*{Acknowledgment}
% The information, data, or work presented herein was funded in part by the 1) U.S. Office of Naval Research under award number N000142212239, and 2) National
% Science Foundation (NSF-AMPS) under award number CON0002619.

\section*{Acknowledgment}
The information, data, or work presented herein was
funded in part by the U.S. Office of Naval Research
under award number N000142212239, and 2) National
Science Foundation (NSF-AMPS) under award number
2229074.

\bibliographystyle{unsrt}
\bibliography{bib.bib}

\begin{thebibliography}{10}

\bibitem{IEA2022}
{IEA} (2022), {S}mart {G}rids.
\newblock \url{https://www.iea.org/reports/smart-grids/}, year = 2022.
\newblock License: CC BY 4.0.

\bibitem{7150418}
Julio Barros, Jose~Julio Gutiérrez, Matilde de~Apráiz, Purificación Saiz, Ramón~I. Diego, and Andoni Lazkano.
\newblock Rapid voltage changes in power system networks and their effect on flicker.
\newblock {\em IEEE Transactions on Power Delivery}, 31(1):262--270, 2016.

\bibitem{8466598}
Kaveh Dehghanpour, Zhaoyu Wang, Jianhui Wang, Yuxuan Yuan, and Fankun Bu.
\newblock A survey on state estimation techniques and challenges in smart distribution systems.
\newblock {\em IEEE Transactions on Smart Grid}, 10(2):2312--2322, 2019.

\bibitem{zhao2019power}
Junbo Zhao, Antonio G{\'o}mez-Exp{\'o}sito, Marcos Netto, Lamine Mili, Ali Abur, Vladimir Terzija, Innocent Kamwa, Bikash Pal, Abhinav~Kumar Singh, Junjian Qi, et~al.
\newblock Power system dynamic state estimation: Motivations, definitions, methodologies, and future work.
\newblock {\em IEEE Transactions on Power Systems}, 34(4):3188--3198, 2019.

\bibitem{zamzam_data-driven_2019}
Ahmed~S. Zamzam, Xiao Fu, and Nicholas~D. Sidiropoulos.
\newblock Data-{Driven} {Learning}-{Based} {Optimization} for {Distribution} {System} {State} {Estimation}, March 2019.
\newblock Issue: arXiv:1807.01671 arXiv:1807.01671 [eess].

\bibitem{azimian_state_2022}
Behrouz Azimian, Reetam~Sen Biswas, Shiva Moshtagh, Anamitra Pal, Lang Tong, and Gautam Dasarathy.
\newblock State and {Topology} {Estimation} for {Unobservable} {Distribution} {Systems} {Using} {Deep} {Neural} {Networks}.
\newblock {\em IEEE Transactions on Instrumentation and Measurement}, 71:1--14, 2022.

\bibitem{vu_cyber-physical_2020}
Tuyen~V. Vu, Bang~L.H. Nguyen, Zheyuan Cheng, Mo-Yuen Chow, and Bin Zhang.
\newblock Cyber-{Physical} {Microgrids}: {Toward} {Future} {Resilient} {Communities}.
\newblock {\em IEEE Industrial Electronics Magazine}, 14(3):4--17, 2020.

\bibitem{zhao_robust_2017}
Junbo Zhao, Marcos Netto, and Lamine Mili.
\newblock A {Robust} {Iterated} {Extended} {Kalman} {Filter} for {Power} {System} {Dynamic} {State} {Estimation}.
\newblock {\em IEEE Transactions on Power Systems}, 32(4):3205--3216, July 2017.

\bibitem{nguyen2021distributed}
Bang~LH Nguyen, Tuyen~V Vu, Joseph~M Guerrero, Mischael Steurer, Karl Schoder, and Tuan Ngo.
\newblock Distributed dynamic state-input estimation for power networks of microgrids and active distribution systems with unknown inputs.
\newblock {\em Electric Power Systems Research}, 201:107510, 2021.

\bibitem{s21062085}
Xue-Bo Jin, Ruben~Jonhson Robert~Jeremiah, Ting-Li Su, Yu-Ting Bai, and Jian-Lei Kong.
\newblock The new trend of state estimation: From model-driven to hybrid-driven methods.
\newblock {\em Sensors}, 21(6), 2021.

\bibitem{9072507}
Ahmed~Samir Zamzam and Nicholas~D. Sidiropoulos.
\newblock Physics-aware neural networks for distribution system state estimation.
\newblock {\em IEEE Transactions on Power Systems}, 35(6):4347--4356, 2020.

\bibitem{9126036}
Lei Wang, Qun Zhou, and Shuangshuang Jin.
\newblock Physics-guided deep learning for power system state estimation.
\newblock {\em Journal of Modern Power Systems and Clean Energy}, 8(4):607--615, 2020.

\bibitem{10143253}
Di~Cao, Junbo Zhao, Weihao Hu, Nanpeng Yu, Jiaxiang Hu, and Zhe Chen.
\newblock Physics-informed graphical learning and bayesian averaging for robust distribution state estimation.
\newblock {\em IEEE Transactions on Power Systems}, pages 1--13, 2023.

\bibitem{10238825}
Tong Su, Junbo Zhao, Yansong Pei, and Fei Ding.
\newblock Probabilistic physics-informed graph convolutional network for active distribution system voltage prediction.
\newblock {\em IEEE Transactions on Power Systems}, 38(6):5969--5972, 2023.

\bibitem{9647978}
Wenyu Wang and Nanpeng Yu.
\newblock Estimate three-phase distribution line parameters with physics-informed graphical learning method.
\newblock {\em IEEE Transactions on Power Systems}, 37(5):3577--3591, 2022.

\bibitem{pagnier2021physics}
Laurent Pagnier and Michael Chertkov.
\newblock Physics-informed graphical neural network for parameter \& state estimations in power systems.
\newblock {\em arXiv preprint arXiv:2102.06349}, 2021.

\bibitem{ostrometzky2019physics}
Jonatan Ostrometzky, Konstantin Berestizshevsky, Andrey Bernstein, and Gil Zussman.
\newblock Physics-informed deep neural network method for limited observability state estimation.
\newblock {\em arXiv preprint arXiv:1910.06401}, 2019.

\bibitem{de2022physics}
Steven de~Jongh, Frederik Gielnik, Felicitas Mueller, Loris Schmit, Michael Suriyah, and Thomas Leibfried.
\newblock Physics-informed geometric deep learning for inference tasks in power systems.
\newblock {\em Electric Power Systems Research}, 211:108362, 2022.

\bibitem{9265470}
Mahdi Khodayar, Guangyi Liu, Jianhui Wang, and Mohammad~E. Khodayar.
\newblock Deep learning in power systems research: A review.
\newblock {\em CSEE Journal of Power and Energy Systems}, 7(2):209--220, 2021.

\bibitem{yarlagadda_power_2021}
R.~Yarlagadda, V.~Kosana, and K.~Teeparthi.
\newblock Power {System} {State} {Estimation} and {Forecasting} using {CNN} based {Hybrid} {Deep} {Learning} {Models}.
\newblock In {\em 2021 {IEEE} {International} {Conference} on {Technology}, {Research}, and {Innovation} for {Betterment} of {Society} ({TRIBES})}, pages 1--6, 2021.

\bibitem{he_power_2020}
Yi~He, Songjian Chai, Zhao Xu, Chun~Sing Lai, and Xu~Xu.
\newblock Power system state estimation using conditional generative adversarial network.
\newblock {\em IET Generation, Transmission \& Distribution}, 14, December 2020.

\bibitem{zhang_real-time_2019}
Liang Zhang, Gang Wang, and Georgios~B. Giannakis.
\newblock Real-time {Power} {System} {State} {Estimation} and {Forecasting} via {Deep} {Neural} {Networks}.
\newblock {\em IEEE Transactions on Signal Processing}, 67(15):4069--4077, August 2019.
\newblock Number: 15 arXiv:1811.06146 [cs, stat].

\bibitem{zhou2020graph}
Jie Zhou, Ganqu Cui, Shengding Hu, Zhengyan Zhang, Cheng Yang, Zhiyuan Liu, Lifeng Wang, Changcheng Li, and Maosong Sun.
\newblock Graph neural networks: A review of methods and applications.
\newblock {\em AI open}, 1:57--81, 2020.

\bibitem{10050763}
Rahul Madbhavi, Balasubramaniam Natarajan, and Babji Srinivasan.
\newblock Graph neural network-based distribution system state estimators.
\newblock {\em IEEE Transactions on Industrial Informatics}, pages 1--10, 2023.

\bibitem{Park2023DistributedPS}
Sangwoo Park, Fernando Gama, Javad Lavaei, and Somayeh Sojoudi.
\newblock Distributed power system state estimation using graph convolutional neural networks.
\newblock In {\em Hawaii International Conference on System Sciences}, 2023.

\bibitem{KUNDACINA2023101056}
Ognjen Kundacina, Mirsad Cosovic, Dragisa Miskovic, and Dejan Vukobratovic.
\newblock Graph neural networks on factor graphs for robust, fast, and scalable linear state estimation with pmus.
\newblock {\em Sustainable Energy, Grids and Networks}, 34:101056, 2023.

\bibitem{nguyen_spatial-temporal_2022}
Bang Nguyen, Tuyen Vu, Thai-Thanh Nguyen, Mayank Panwar, and Rob Hovsapian.
\newblock Spatial-temporal recurrent graph neural networks for fault diagnostics in power distribution systems.
\newblock {\em arXiv preprint arXiv:2210.15177}, 2022.

\bibitem{9429985}
Laura von Rueden, Sebastian Mayer, Katharina Beckh, Bogdan Georgiev, Sven Giesselbach, Raoul Heese, Birgit Kirsch, Julius Pfrommer, Annika Pick, Rajkumar Ramamurthy, Michal Walczak, Jochen Garcke, Christian Bauckhage, and Jannis Schuecker.
\newblock Informed machine learning – a taxonomy and survey of integrating prior knowledge into learning systems.
\newblock {\em IEEE Transactions on Knowledge and Data Engineering}, 35(1):614--633, 2023.

\bibitem{RAISSI2019686}
M.~Raissi, P.~Perdikaris, and G.E. Karniadakis.
\newblock Physics-informed neural networks: A deep learning framework for solving forward and inverse problems involving nonlinear partial differential equations.
\newblock {\em Journal of Computational Physics}, 378:686--707, 2019.

\bibitem{misyris2020physics}
George~S Misyris, Andreas Venzke, and Spyros Chatzivasileiadis.
\newblock Physics-informed neural networks for power systems.
\newblock In {\em 2020 IEEE Power \& Energy Society General Meeting (PESGM)}, pages 1--5. IEEE, 2020.

\bibitem{9743327}
Bin Huang and Jianhui Wang.
\newblock Applications of physics-informed neural networks in power systems - a review.
\newblock {\em IEEE Transactions on Power Systems}, 38(1):572--588, 2023.

\bibitem{ji2021stiff}
Weiqi Ji, Weilun Qiu, Zhiyu Shi, Shaowu Pan, and Sili Deng.
\newblock Stiff-pinn: Physics-informed neural network for stiff chemical kinetics.
\newblock {\em The Journal of Physical Chemistry A}, 125(36):8098--8106, 2021.

\bibitem{9779551}
Shuai Zhao, Yingzhou Peng, Yi~Zhang, and Huai Wang.
\newblock Parameter estimation of power electronic converters with physics-informed machine learning.
\newblock {\em IEEE Transactions on Power Electronics}, 37(10):11567--11578, 2022.

\bibitem{9654642}
Md~Jakir Hossain and Mahshid Rahnamay–Naeini.
\newblock State estimation in smart grids using temporal graph convolution networks.
\newblock In {\em 2021 North American Power Symposium (NAPS)}, pages 01--05, 2021.

\bibitem{kipf2017semisupervised}
Thomas~N. Kipf and Max Welling.
\newblock Semi-supervised classification with graph convolutional networks.
\newblock In {\em International Conference on Learning Representations}, 2017.

\bibitem{9395439}
Nurul~A. Asif, Yeahia Sarker, Ripon~K. Chakrabortty, Michael~J. Ryan, Md.~Hafiz Ahamed, Dip~K. Saha, Faisal~R. Badal, Sajal~K. Das, Md.~Firoz Ali, Sumaya~I. Moyeen, Md.~Robiul Islam, and Zinat Tasneem.
\newblock Graph neural network: A comprehensive review on non-euclidean space.
\newblock {\em IEEE Access}, 9:60588--60606, 2021.

\bibitem{9064519}
Rahul Rai and Chandan~K. Sahu.
\newblock Driven by data or derived through physics? a review of hybrid physics guided machine learning techniques with cyber-physical system (cps) focus.
\newblock {\em IEEE Access}, 8:71050--71073, 2020.

\bibitem{MOHAMMADIAN2023109551}
Mostafa Mohammadian, Kyri Baker, and Ferdinando Fioretto.
\newblock Gradient-enhanced physics-informed neural networks for power systems operational support.
\newblock {\em Electric Power Systems Research}, 223:109551, 2023.

\bibitem{liao2021review}
Wenlong Liao, Birgitte Bak-Jensen, Jayakrishnan~Radhakrishna Pillai, Yuelong Wang, and Yusen Wang.
\newblock A review of graph neural networks and their applications in power systems.
\newblock {\em Journal of Modern Power Systems and Clean Energy}, 10(2):345--360, 2021.

\bibitem{tian_neural-network-based_2021}
Guanyu Tian, Yingzhong Gu, Di~Shi, Jing Fu, Zhe Yu, and Qun Zhou.
\newblock Neural-network-based {Power} {System} {State} {Estimation} with {Extended} {Observability}.
\newblock {\em Journal of Modern Power Systems and Clean Energy}, 9(5):1043--1053, September 2021.

\bibitem{labach2019survey}
Alex Labach, Hojjat Salehinejad, and Shahrokh Valaee.
\newblock Survey of dropout methods for deep neural networks.
\newblock {\em arXiv preprint arXiv:1904.13310}, 2019.

\bibitem{masters2018revisiting}
Dominic Masters and Carlo Luschi.
\newblock Revisiting small batch training for deep neural networks.
\newblock {\em arXiv preprint arXiv:1804.07612}, 2018.

\bibitem{fey2019fast}
Matthias Fey and Jan~Eric Lenssen.
\newblock Fast graph representation learning with pytorch geometric.
\newblock {\em arXiv preprint arXiv:1903.02428}, 2019.

\bibitem{8962079}
Zhiyuan Cao, Yubo Wang, Chi-Cheng Chu, and Rajit Gadh.
\newblock Scalable distribution systems state estimation using long short-term memory networks as surrogates.
\newblock {\em IEEE Access}, 8:23359--23368, 2020.

\end{thebibliography}

% \vspace{12pt}
% \color{red}
% IEEE conference templates contain guidance text for composing and formatting conference papers. Please ensure that all template text is removed from your conference paper prior to submission to the conference. Failure to remove the template text from your paper may result in your paper not being published.

\end{document}